\documentclass[12pt]{article}
\usepackage[margin=1.2in]{geometry}
\usepackage[utf8]{inputenc}
\usepackage[english]{babel}
\usepackage{amsmath,amssymb,latexsym}
\usepackage{slashed}
\usepackage{graphicx}
\DeclareGraphicsExtensions{.png}
\usepackage{epstopdf}
\usepackage{relsize}
\usepackage{mathtools}
\usepackage[stdtext]{youngtab}
\usepackage{subcaption}
\usepackage{csquotes}
\usepackage[sorting=none, backend=bibtex]{biblatex}
\usepackage{tikz}
\usetikzlibrary{shapes.gates.logic.US, decorations.markings,decorations.pathreplacing,calc}
\usepackage{array,booktabs}
\usepackage{lmodern}
\usepackage{microtype}
\usepackage{xcolor}
\usepackage{hyperref}
\hypersetup{
  colorlinks=false,
  urlbordercolor=gray,
  citebordercolor=gray,
  linkbordercolor=gray,
  pdfborderstyle={/S/U/W 1}
}
\AtBeginDocument{%
  }
\usepackage{pict2e}

\def \sec{\begin{section}}
\def \esec{\end{section}}
\def \beq {\begin{equation}}
\def \eeq {\end{equation}}

\def \la {\lambda}

\def \si {\sigma}

\def \Ga {\Gamma}

\def \ep {\epsilon}

\def \Si {\Sigma}
\def \si {\sigma}

\def \Wc {\mathcal{W}}
\def \Rc {\mathcal{R}}
\def \RR {\mathbb{R}}

\def \ra {\rightarrow}

\DeclareMathOperator*{\res}{res}

\DeclareMathOperator\Tr{Tr}

\DeclareMathOperator\diag{diag}

\def \l {\left(}
\def \r {\right)}
\def \ll {\left \langle}
\def \rr {\right \rangle}

\def \wc {\mathcal{W}}

\newcommand{\Nsusy}{{\mathcal{N}}}
\makeatletter
\newcommand{\prodprime}{\mathop{\mathpalette{\prodprime@}{\prod}}}
\newcommand{\prodprime@}[2]{\smash{\sideset{}{'}{#1#2}}\vphantom{#1#2}}
\makeatother
\newcommand{\compressthisequation}{%
  \divide\thinmuskip2\relax
  \divide\medmuskip2\relax
  \divide\thickmuskip2\relax}

\bibliography{ref}
\begin{document}

\begin{center}

{  \Large \bf Surface defects and instanton--vortex interaction }

\end{center}

\vspace{1mm}

\begin{center}
    
{\large
    
  A.~Gorsky$^{\,3,4}$, B.~Le~Floch$^{\,2}$, A.~Milekhin$^{\,1,3,5}$\ and N.~Sopenko$^{\,4,5}$\ }

\vspace{3mm}

$^1$ Department of Physics, Princeton University, Princeton, NJ 08544, USA \\
$^2$ Princeton Center for Theoretical Science, Princeton University, Princeton, NJ 08544, USA \\
$^3$ Institute for Information Transmission Problems of Russian Academy of Science, B. Karetnyi 19, Moscow 127051, Russia\\
$^4$ Moscow Institute of Physics and Technology, Dolgoprudny 141700, Russia \\
$^5$ Institute of Theoretical and Experimental Physics, B.Cheryomushkinskaya 25, Moscow 117218, Russia \\

\vspace{1cm}

gorsky@itep.ru, blefloch@princeton.edu, milekhin@princeton.edu, niksopenko@gmail.com

\end{center}

\vspace{1cm}

\begin{center}

{\large \bf Abstract}
\end{center}
We propose a simple formula for the 4d--2d partition function of half-BPS surface defects in $d=4$, $\mathcal{N}=2$ gauge theories: $Z^{\text{4d--2d}}=\langle Z^{\text{2d}} \rangle_{\text{4d}}$.
Our results are applicable for any surface defect obtained by gauging a 2d~flavour symmetry using a 4d~gauge group.
For defects obtained via the Higgsing procedure, our formula reproduces the recent calculation by Pan and Peelaers.
For Gukov--Witten defects our results reproduce the orbifold calculation by Kanno and Tachikawa.
We emphasize the role of ``negative vortices'' which are realized as negative D0 branes.

\newpage

\tableofcontents

\section{Introduction}

The interplay between two-dimensional theories on various defects and the field theory
in the 4d bulk is the subject of intensive studies  focused on evaluating the corresponding
partition functions and relations between them. To some extent these can be thought of as self-consistency
conditions between 2d and 4d theories that allow to treat the defect as a probe which does not destroy the bulk
theory. The 2d/4d duality was first analyzed in \cite{dorey98} in the context of BPS counting, however
more refined mappings between the two systems have later been considered. In particular the 2d
superpotential on the defect can be related to its 4d counterpart \cite{surf}. The most interesting
issue concerns the interplay between 2d and 4d non-perturbative effects that influence the
total 4d--2d partition function. In particular there is the interesting appearance of the
4d gluino condensate in the defect theory \cite{shifman} which is consistent with the
analysis in \cite{surf}.

The analysis becomes more rigid in the $\Omega$-background since we can follow the dependence
on parameters of the deformation. In the deformed case we have new counting parameters
and several interesting limits. The defect corresponding to the surface operators can be introduced
into the theory as a boundary condition, following Gukov--Witten \cite{geolang}, or via the Higgsing
procedure \cite{razamat}.  From the string theory viewpoint, the Higgsing procedure is a reincarnation
of the well-known geometric transition \cite{gopa}.

Recently, the so-called negative branes have been studied \cite{negative} and it was argued that these
are natural objects to get worldvolume theories with a gauge supergroup. An interesting question is
whether these negative branes could play the role of non-perturbative solutions
in ordinary gauge theories with conventional gauge groups. We find that indeed these negative branes
have to be taken into account in the total 4d--2d partition function. Understanding the role
of negative branes is one of the main motivations of our study.

The aim of our study is to get a compact expression for the partition function of 4d
$\mathcal{N}=2$ supersymmetric gauge theories in the Omega-background coupled to 2d
$\mathcal{N}=(2,2)$ linear quiver gauge theory by gauging 2d flavor symmetries using
4d vector multiplets. In this construction the 4d Higgs VEV $\ll i \Phi\rr =i \Si$ coincides with 2d twisted masses.
Purely 2d partition functions have a Coulomb branch representation as a contour integral over the 2d vector
multiplet Higgs field $i\si$ of a perturbative result \cite{Benini:2012ui,2dvortex} (see \autoref{b:2d} for a brief review):
\beq
Z^{\text{2d}}_{\text{vortex}} = \int d \si\, Z^{\text{2d}}(i\si,i\Si)
\eeq
We have found that the inclusion of the 4d degrees of freedom can be obtained by
averaging over the coupling with 4d degrees of freedom.  The main result of this
paper is
\beq
\label{mainresult}
Z^{\text{4d--2d}}_{\text{instanton--vortex}} = \int d\si\, \ll Z^{\text{2d}}(i\si,i\Phi) \rr_{\text{4d}}
\eeq
which is a direct generalization of the Gaiotto--Gukov--Seiberg prescription for the
4d--2d twisted superpotential \cite{surf}:
\beq
\wc^{\text{4d--2d}} = \ll\wc^{\text{2d}}(i\Phi) \rr_{\text{4d}} \,.
\eeq
Later we will explain how to compute the 4d VEV in (\ref{mainresult}) and how to choose the integration contour.
One can compute the contour integral over $\si$ and obtain the 4d--2d instanton--vortex
partition function\footnote{For intersecting defects we conjecture that the full partition function is also given by the 4d VEV: $Z^{\text{4d--2d--2d'--0d}}_{\text{instanton--vortex}}=\sum_{\la,k,k'} Z^{
\text{4d}}(\la) Z^{\text{4d--2d}}(\la,k)Z^{\text{4d--2d'}}(\la,k') Z^{\text{2d}}(k)Z^{\text{2d'}}(k') Z^\text{0d}(k,k')$. }
\beq
\label{introZiv}
  Z^{\text{4d--2d}}_{\text{instanton--vortex}}=\sum_{\la,k} Z^{\text{4d}}(\la) Z^{\text{4d--2d}}(\la,k) Z^{\text{2d}}(k)
\eeq
where the sum goes over instantons $\la$ and vortex
numbers $k$.  While the 2d theory in isolation only involves vorticities
$k\geq0$, the interaction with instantons allows negative~$k$.  We
interpret this as fractionalization of an instanton into a vortex and an
``instanton--negative-vortex''.
Besides, apart from the standard instanton configurations, $\la$ includes instantons trapped on the surface defect.

Here, $Z^{\text{4d}}(\la)$ is the usual instanton
contribution found by Nekrasov and $Z^{\text{2d}}(k)$ is
a standard 2d vortex contribution, in the zero-instanton
sector.  Finally, $Z^{\text{4d--2d}}(\la,k)$
describes the interaction of instantons and vortices.  It vanishes when
there are no instantons:
\beq
Z^{\text{4d--2d}}(\emptyset,k)=1 \,.
\eeq
However it is non-trivial even in the absence of vortices, since we still have to take into account the 2d perturbative
contribution in the instanton background.

\begin{figure}
\centering
\begin{tikzpicture}[brane/.style={line width=1pt}, d4/.style={line width=2.4pt}]
  \draw[brane] (0,0) -- (0,3.7) node [below left] {NS5};
  \draw[brane] (3,0) -- (3,3.7);
  \draw[d4] (0,1) -- (3,1);
  \draw[d4] (0,1.8) -- (3,1.8);
  \draw[d4] (0,3) -- (3,3) node [midway, above left] {D4};
  \draw[densely dashed,{stealth}-{stealth},thick] (0.03,.3) -- (2.97,.3) node [midway, below] {$x^4$};
  \draw[densely dashed,{stealth}-{stealth},thick] (3.1,1) -- (3.1,1.77) node [midway, right] {$a_2-a_3$};
  \draw[densely dashed,{stealth}-{stealth},thick] (3.1,1.83) -- (3.1,3) node [midway, right] {$a_1-a_2$};
\end{tikzpicture}
\caption{$U(3)$ pure gauge theory. The distance $x^4=1/g^2_{\text{4d}}$ between NS5 branes in the 4-direction is the 
instanton counting parameter.
Coordinates of D4 branes in the $(x^5,x^6)$ plane set the corresponding Higgs VEV.}
\label{d4_un}
\end{figure}
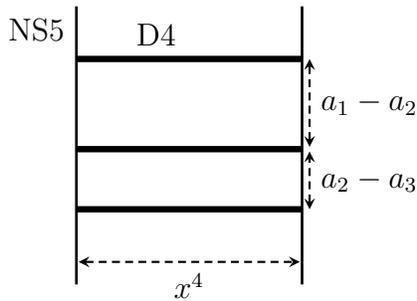

\begin{figure}
\centering
\begin{tikzpicture}[brane/.style={line width=1pt}, d4/.style={line width=2.4pt}]
  \begin{scope}
    \draw[brane] (0,0) -- (0,2.7) node [below left] {NS5};
    \draw[brane] (3,0) -- (3,2.7);
    \draw[d4] (0,1) -- (3,1);
    \draw[brane] (1.4,0.9) -- (3,0.9) node [near end, outer sep = 9pt, anchor = west ] {D0} ;
    \draw[brane,red] (-0.5,-1) -- (1.5,1) node [midway, below right] {\textcolor{black}{D2}} ;
    \draw[d4] (0,1.6) -- (3,1.6);
    \draw[d4] (0,2) -- (3,2) node [midway, above left] {D4};
    \draw[brane] (-0.5,-1.5) -- (-0.5,0);
  \end{scope}
  \node at (5,1) {\Huge$\simeq$};
  \begin{scope}[shift={(7,0)}]
    \draw[brane] (0,0) -- (0,2.7) node [below left] {NS5};
    \draw[brane] (3,0) -- (3,2.7);
    \draw[d4] (0,1) -- (3,1);
    \draw[brane] (0,0.9) -- (3,0.9) node [near end, outer sep = 20pt, anchor = west ] {D0} ;
    \draw[brane,blue] (0,0.8) -- (1.3,0.8) node [near start, outer sep = 9pt, anchor = east ] {$D0_n$} ;
    \draw[brane,red] (-0.5,-1) -- (1.5,1) node [midway, below right] {\textcolor{black}{D2}} ;
    \draw[d4] (0,1.6) -- (3,1.6);
    \draw[d4] (0,2) -- (3,2) node [midway, above left] {D4};
    \draw[brane] (-0.5,-1.5) -- (-0.5,0);
  \end{scope}
\end{tikzpicture}
\caption{$U(3)$ pure gauge theory with a minimal surface operator. The distance in the $x^4$-direction between the left NS5
brane and the D2 brane is the 2D FI parameter. D0 branes between the left NS5 and D2 are essentially two-dimensional
vortices (not shown). We interpret D0 branes between the right NS5 and D2 as a composite object consisting of a long
D0 brane (4D instanton) and a negative vortex.}
\label{d4_def}
\end{figure}
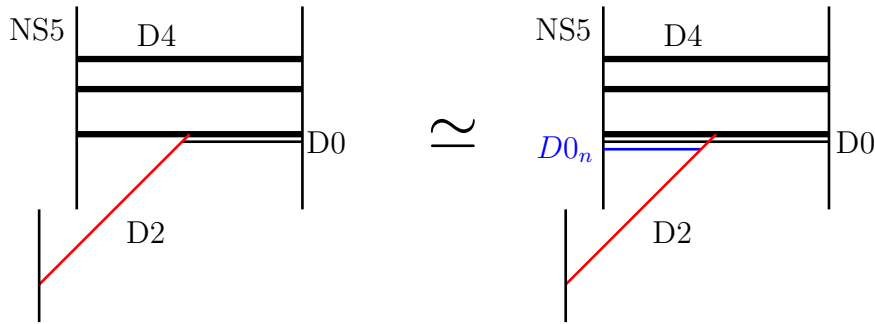

When evaluating the instanton--vortex partition function we will meet contributions
that can be naturally attributed to negative vortices in a given instanton
background. In theories with a brane construction, they can be identified with negative D0 branes. 
Geometrically, 4d $\Nsusy=2$ $U(N)$ pure gauge super Yang--Mills (SYM) theory can be 
obtained by suspending $N$ D4 branes between two NS5 (see \autoref{d4_un}).
Branes are stretched along the following directions:
\begin{center}
\begin{tabular}{rl}\toprule
  NS5: & 0 1 2 3 \phantom{4} 5 6 \\
  D4: & 0 1 2 3 4 \\\bottomrule
\end{tabular}
\end{center}
Instantons are represented by D0 branes stretched along $x^4$ between two NS5.
The simplest surface defect can be engineered by adding a D2 brane along $(x^0,x^1,x^7)$, ending on an additional NS5 brane. The corresponding
2d $\Nsusy=(2,2)$ theory is given by a $U(1)$ vector multiplet coupled to $N$ fundamental chirals.
The surface defect divides the D0 worldline into two pieces. One piece of the D0
brane can be identified as a vortex in the 2d theory on the defect, and the second
piece as an instanton together with a negative vortex of the worldsheet defect
theory (see \autoref{d4_def}).  Note, however, that negative D0 branes are not independent objects: we can add them
only if we already have a ``positive'' D0 brane. In other words, negative vortices appear only in the instanton background.
A negative vortex contribution equals to a vortex contribution inverted. Since standard vortices obey bosonic statistics
(upon a lift to 3d, \cite{dgh}), negative vortices are fermionic. This agrees with the negative D0 brane interpretation
\cite{negative}.  For Gukov--Witten defects fractional instantons have been extensively studied in mathematical literature
under the name of ``ramified instantons'' \cite{kronheimer1993,kronheimer2011}. In this paper we give a concrete physical
realization of ramified instantons as composite objects consisting of usual instantons and a number of (negative) vortices.
The picture has a lot in common with the
fractionalization of an instanton into the composite of instanton--monopoles and
instanton--KK-monopoles \cite{davies} in the SYM theory on $\mathbb{R}^3\times S^1$.
In our case, instanton--negative-vortices play a role similar to the instanton--KK-monopoles.
We will discuss later in \autoref{sec:negativebranes} the nature of these objects.

In the body of the paper we perform several checks of our proposal.

When the defect is derived via Higgsing, we compare with the recent results obtained
in \cite{pp} for intersecting defects on the squashed sphere.  We extend the
analysis of \cite{pp} by working out the case where the Higgsed $SU(N)\times SU(N)$
is a pair of gauge groups, rather than one gauge group and one flavour group.  We
restrict ourselves however to a single surface operator: as we will see the Higgsed
instanton partition function otherwise involves higher-order poles.
The sum takes the form~\eqref{introZiv} for the 4d--2d system depicted in
\autoref{fig:HiggsSUNSUN}.

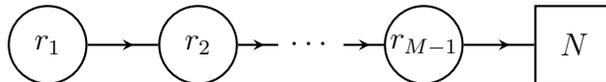
\begin{figure}
\centering
\begin{tikzpicture}[thick,
  ->-/.style 2 args={decoration={markings, mark=at position #1 with {\arrow[rotate=#2]{stealth}}},postaction={decorate}}]
  \node[circle,draw,minimum size=2.5em] (1) at (0,0) {$r_1$};
  \node[circle,draw,minimum size=2.5em] (2) at (2,0) {$r_2$};
  \node (dots) at (3.5,0) {$\cdots$};
  \node[circle,draw,minimum size=2.5em] (M1) at (5,0) {$\mathclap{r_{M-1}}$};
  \node[draw,minimum size=2.5em] (N) at (7,0) {$N$};
  \draw[->-={.65}{0}] (1) -- (2);
  \draw[->-={.65}{0}] (2) -- (dots);
  \draw[->-={.65}{0}] (dots) -- (M1);
  \draw[->-={.65}{0}] (M1) -- (N);
\end{tikzpicture}
\caption{2d quiver diagram corresponding to breaking $U(N) \ra U(n_1)\times \dots \times U(n_M)$ where $r_I = n_1+\cdots+n_I$.  Circles represent gauge factors, squares represent matter flavour symmetries. Arrows represent chiral multiplets that transform in the antifundamental/fundamental representation of the groups at the source/tip of the arrow.}
\label{defect_quiver}
\end{figure}

The second check concerns the Gukov--Witten (monodromy) defect in pure $SU(N)$ SYM,
also described by a 2d $\Nsusy=(2,2)$ quiver gauge theory on the defect.
Breaking $U(N) \to U(n_1) \times \dots \times U(n_M)$ can be achieved by coupling 4d
degrees of freedom to the 2d $\Nsusy=(2,2)$ quiver gauge theory in \autoref{defect_quiver}.
We prove that the instanton--vortex partition function given by our prescription coincides with the orbifold calculation of Kanno--Tachikawa \cite{orbi}.

The paper is organized as follows.  In \autoref{sec:coupling4dto2d} we write down
our prescription for the Omega-background partition function of 4d--2d coupled gauge
theories. In \autoref{sec:Higgs} we test our
conjecture for defects that arise from Higgsing and make a comment on 3d/5d case.
In \autoref{sec:negativebranes} we
emphasize the importance of instanton fragmentation and instanton--negative-vortices. In
\autoref{sec:orbifold} we compare our conjecture with the orbifold calculation.
Details concerning the Nekrasov partition function and 2d $\Nsusy=(2,2)$ partition
functions are collected in \autoref{a:nek} and \autoref{b:2d}.

\section{Coupling 2d to 4d}\label{sec:coupling4dto2d}

\subsection{Coupling 2d free chirals}

Let us start with a simple example of a surface operator in 4d $\Nsusy=2$ $U(N)$
SYM.  We gauge the $U(N)$ flavor symmetry of $N$ free 2d $\Nsusy=(2,2)$ chiral
multiplets using the 4d vector multiplet restricted to the surface.  The 4d Higgs
field $i \Phi$ has VEV
$\ll i \Phi\rr=i\Si =\diag(i\Si_1,\ldots,i\Si_N)$ and the
$i\Sigma_A$ play the role of 2d twisted masses. We will denote the overall 2d $U(1)$
factor by $i\si$. Note that in our conventions all the scalar fields have $i$ in
front (if one considers the theories on spheres this makes $\Sigma$ and $\sigma$
real).  The general strategy of 4d instanton calculations is to describe the
instanton moduli spaces using ADHM equations and afterwards include all additional
contributions (matter fields, for instance) as some characteristic classes which one
integrates over the ADHM moduli space.  Here we use exactly the same strategy.

Let us first review how the 4d $\Nsusy=2$ fundamental hypermultiplet contribution is
derived.  To do it we should first understand how to compute VEVs of various chiral
operators.  The standard Nekrasov partition function for the pure gauge $U(N)$
SYM theory can be written as a sum over $N$ Young diagrams $\la_A$, $A=1,\ldots,N$
\beq
\label{znek}
Z^{\text{4d}\ U(N)\ \text{SYM}}_{\text{instanton}}=\sum_{\la} q^{|\la|} Z^{\text{4d}}_{\text{vector}}(i\Sigma,\la)
\eeq

The precise form of $Z^{\text{4d}}_{\text{vector}}$ can be found in \autoref{a:nek},
but for now it is not important.  The instanton number
$K=|\la|=|\la_1|+\ldots+|\la_N|$ is equal to the total number of boxes.  It is
convenient to parametrize all the boxes using an index $I=(A,r,s)$, where $A$
selects a Young diagram $\la_A$ and $(r,s)\in\la_A$ with $r,s \geq 1$ is a box in
$\la_A$.  The so-called ``box content'' is
\beq
\label{phii}
\phi_I \equiv \phi_{(A,r,s)} = i\Si_A + (r-\tfrac{1}{2}) \ep_1 + (s-\tfrac{1}{2}) \ep_2 \,.
\eeq

Actually, one can take a step back and consider the contour integral representation
of the partition function (\ref{znek}).  In this case the $K$-instanton contribution
is given by the contour integral over $K$ variables $\phi_I$, and (\ref{phii})
defines positions of the poles.

Suppose we want to compute the VEV of a single-trace\footnote{We denote the trace
  in the fundamental representation by $\Tr \equiv \Tr_\square$.} chiral operator
$\Tr f(i\Phi)$ inserted at the origin\footnote{The general Omega-deformation
  preserves only one supercharge $\tilde{Q}$. Using the explicit form of the
  supersymmetry algebra it is easy to show that $\tilde{Q}$ annihilates $\Phi$ at
  the origin of $\RR^4$.  Or, in the case of a squashed $S^4$, at the north pole
  (and $\bar{\Phi}$ at the south pole).
  Thus $\Phi(0)$ is a good $\tilde{Q}$-closed observable. We refer to
  \cite{NekOk} for details.} of $\mathbb{R}^4$. Then the partition function
(\ref{znek}) is turned into \cite{small_little}
\beq
\label{chiral}
\ll \Tr f(i\Phi) \rr=\sum_{\{\la\}}q^{|\la|}  Z^{\text{4d}}_{\text{vector}}(i\Sigma,\la) \ll \Tr f(i\Phi) \rr_\la
\eeq
where $\ll \Tr f(i\Phi) \rr_\la$ is defined as:
\beq
\label{formula}
\ll \Tr f(i\Phi) \rr_\la = \sum^N_{A=1} f(i\Si_A) -
 \sum_{I=1}^K \begin{multlined}[t]
   \bigl( f(\phi_I-\tfrac{\ep_1+\ep_2}{2})-f(\phi_I+\tfrac{\ep_1-\ep_2}{2})\\
   -f(\phi_I-\tfrac{\ep_1-\ep_2}{2})+f(\phi_I+\tfrac{\ep_1+\ep_2}{2}) \bigr) \,.
 \end{multlined}
\eeq

Obviously the first term is the perturbative contribution and the sum over $I$ is
purely an instanton contribution.  For multi-trace operators we use the
factorization property
\beq
\ll \Tr f_1(i\Phi) \Tr f_2(i\Phi) \rr_\la = \ll \Tr f_1(i\Phi) \rr_\la \ll \Tr f_2(i\Phi) \rr_\la \,.
\eeq
Let us now discuss various 4d matter contributions.
Without the Omega-background, the perturbative contribution to the 4d prepotential
from a 4d hypermultiplet of mass $m$ has a standard 1-loop Coleman--Weinberg
form\footnote{Here $i\Si-m$ means $\diag(i\Si_1-m,\ldots,i\Si_N-m)$.}
\beq
\label{mpert}
\cfrac{3}{4} \Tr(i\Si-m)^2 - \cfrac{1}{2} \Tr(i\Si-m)^2 \log(i\Si-m) =  \int_0^\infty \cfrac{ds}{s^3} \Tr e^{-s(i\Si-m)} \,.
\eeq
If we switch on the Omega-deformation then instead of (\ref{mpert}) we will have a 4d Schwinger-like answer \cite{mtop1,mtop2}:
\beq
Z^{\text{4d}\ U(N),\square}_{\text{instanton}} = \ll \exp \l \int_0^\infty \cfrac{ds}{s} \cfrac{\Tr e^{-s(i\Phi-m)}}{(1-e^{-s \ep_1})(1-e^{-s \ep_2}) } \r \rr \,.
\eeq
The previous formula is exactly the 4d Nekrasov partition function of a $U(N)$ gauge
theory with a hypermultiplet in the fundamental representation.  Indeed, using
(\ref{chiral}) it is easy too see that it is the product of a perturbative
contribution and an instanton contribution.  The perturbative contribution
\beq
\label{hyper_pert}
\exp \l \int_0^\infty \cfrac{ds}{s} \cfrac{\Tr e^{-s(i\Si-m)}}{(1-e^{-s \ep_1})(1-e^{-s \ep_2}) } \r
\eeq
reduces in the limit $\ep_1,\ep_2 \ra 0$ to the perturbative contribution
(\ref{mpert}) divided by $\ep_1 \ep_2$, as expected.  The instanton contribution
reads
\beq
\label{hyper_inst}
Z^{\text{4d}}_{\square}(m,i\Si,\la) = \prod_I \l \phi_I - m-\frac{\ep_1+\ep_2}{2} \r \,.
\eeq

For a hypermultiplet in a general representation $\Rc$ one can use the following formula:
\beq
Z^{\text{4d}\ U(N),\Rc}_{\text{instanton}} = \ll \exp \l
\int_0^\infty \cfrac{ds}{s} \cfrac{\Tr_\Rc e^{-s(i\Phi-m)}}{(1-e^{-s \ep_1})(1-e^{-s \ep_2}) } \r  \rr \,.
\eeq
However, one should keep in mind that the 4d gauge group is $U(N)$, not $SU(N)$. For example, for
the antifundamental representation $\overline{\square}$
one will have
\beq
\ll \Tr_{\overline{\square}} e^{i \Phi} \rr_\la = e^{-i \Si_A} - (1-e^{-\ep_1})(1-e^{-\ep_2}) \sum_I e^{-\phi_I+\frac{\ep_1+\ep_2}{2}}
\eeq
and for the adjoint representation,
\beq
\ll \Tr_{\text{adj}} e^{i \Phi} \rr_\la = \ll \Tr_{\overline{\square}} e^{i \Phi} \rr_\la \ll \Tr_{\square} e^{i \Phi} \rr_\la \,.
\eeq
In the next subsection we will explain how to write down a compact formula for a arbitrary representation using supersymmetric
Schur polynomials.

Let us comment on why this formula works so well.  It will be convenient to discuss
the $S^4$ case.  In the case of $S^4$ localization \cite{pestun}, the saddle-point
configuration of the Higgs field $i \Phi$ is constant, the gauge connection is zero
and scalars of the hypermultiplet are zero.  This is why we obtain the standard
text-book Schwinger answer (\ref{hyper_pert}) for the hypermultiplet
determinant. Instanton corrections in case of $S^4$ arise as point-like instantons
at the north and south poles.  Exactly at these points we can use (\ref{formula}) to
take quantum fluctuations into account.

Let us now consider a 2d chiral multiplet in the fundamental representation of a
4d gauge group and the antifundamental of a 2d gauge group.
Without coupling to 4d and to the Omega-background we have
a standard 1-loop 2d Coleman--Weinberg answer for the 2d twisted superpotential (in
our conventions $i\si$ is the value of the 2d vector multiplet scalar):
\beq
\Wc^{\text{2d}}=\Tr(i\si-i \Si) \log(i\si-i\Si) - \Tr(i\si-i \Si)  = -\int_0^\infty \cfrac{ds}{s^2} \Tr e^{-s(i\Si -i\si)} \,.
\eeq
Now let us switch on the Omega deformation and couple to 4d.  We place the surface
defect in the $\ep_2$ plane.  We propose the following answer for the full 4d--2d
partition function\footnote{The shift by $\ep_2$ in the exponent is a matter of
  convention.}:
\beq
\label{4d2d}
\ll \exp \l  - \int_0^\infty \cfrac{ds}{s} \cfrac{\Tr e^{-s(i\Phi-i\si+\ep_2)}}{1-e^{-s \ep_2}} \r \rr_{\text{4d}} \,.
\eeq
This expression is not new: it was obtained independently in \cite{kim} using a 
Higgsing procedure, in \cite{koro} using Chern characters and in \cite{torus} 
using topological vertex techniques.

In the limit $\ep_1,\ep_2 \ra 0$, the sum over Young diagrams in (\ref{znek}) localizes on 
the saddle-point configuration $\la^*$ \cite{NekOk} and it is not difficult to see
that in this limit (\ref{4d2d}) leads to
\beq
\exp \l -\cfrac{1}{\ep_2} \ll \Wc^{\text{2d}}  \rr_{\text{4d}} \r
\eeq
This is exactly the Gaiotto--Gukov--Seiberg formula for the 4d--2d twisted
superpotential.

As before, the contribution (\ref{4d2d}) contains the 2d Schwinger-like
perturbative part coming from the chiral multiplet determinant:
\begin{multline}
Z^{\text{2d}}_{\text{chiral}}=\exp \l  - \int_0^\infty \cfrac{ds}{s} \cfrac{\Tr e^{-s(i\Si-i\si+\ep_2)}}{1-e^{-s \ep_2}} \r \\
= \prod_{A=1}^N \prod_{j=0}^{+\infty} \l i\Si_A - i\si + \ep_2 + j \ep_2 \r  \sim \prod_{A=1}^N 
\cfrac{1}{\Ga(1+(i\Si_A-i\si)/\ep_2)} \,.
\end{multline}
One can obtain this expression by introducing a cutoff near $s=0$ and expanding
$1/(1-e^{-s\ep_2})=\sum_{j\geq 0}e^{-js\ep_2}$.  The integral also has a discretized
version that corresponds to the 5d answer on $\RR^4 \times S^1_\beta$ (sending the
radius $\beta \ra 0$ reproduces the integral)
\begin{multline}
-\sum_{n=1}^\infty \cfrac{\Tr e^{-\beta n(i\Si-i\si-\ep_2)}}{n(1-e^{-n \beta \ep_2})} =
-\sum_{n=1}^\infty \sum_{j=0}^{+\infty} \cfrac{ \Tr e^{-\beta n(i\Si-i\si-\ep_2)} e^{-n j \beta \ep_2} }{n} \\
= \sum_{j=0}^{+\infty} \Tr \log(1-e^{-\beta(i\Si - i\si -\ep_2-\ep_2 j)}) \,.
\end{multline}
The contribution (\ref{4d2d}) also contains the 4d--2d interaction
\beq\label{4d--2d-fundamental-interaction}
Z^{\text{4d--2d}}=\prod_{I=1}^k  \frac{\phi_I-i\si+\frac{\ep_1+\ep_2}{2}}{\phi_I-i\si+\frac{\ep_1+\ep_2}{2}-\epsilon_1} \,.
\eeq
The main claim of this paper is that this piece is the only thing that mixes 2d and
4d degrees of freedom.  Therefore, if one gauges the overall 2d $U(1)$ symmetry by
adding a 2d vector multiplet, the full 4d--2d partition function will read as
follows:
\beq
Z^{\text{4d--2d}}_{\text{instanton--vortex}} = \sum_\la \int\! d\si \ q^{|\la|} Z^{\text{4d}}_{\text{vector}}(i\Si,\phi) \ z^{i\sigma/\epsilon_2} Z^{\text{2d}}_{\text{vector}}(i\si) \ Z^{\text{2d}}_{\text{chiral}}(i\Si,i\si) Z^{\text{4d--2d}}(\phi,i\si)
\eeq
Recently a very similar formula was derived by Pan and Peelaers \cite{pp} for
defects obtained by Higgsing.  In the next section we write the general formula for
2d chiral matter in arbitrary representations of 2d and 4d gauge groups.  The
formula covers all possible 2d defects obtained by gauging 2d flavor symmetries.

\subsection{General prescription}
\label{sec:generalprescription}

We will now write our conjecture for the instanton--vortex partition function of any 4d--2d system where the 4d instanton partition function is known, the theories are coupled by gauging 2d flavor symmetries using 4d vector multiplets, and only simple poles contribute (see \autoref{foot:2dsimplepoles}).

Since 4d masses and 2d twisted masses are simply constant background vector multiplet scalars for the flavor symmetries, we omit them from the notations.  Namely, we include them on the same footing as~$\Sigma$ and $\sigma$ respectively.  On the Omega background, the 2d vector R-charge~$r$ of a chiral multiplet combines with its twisted mass~$m$ into a dimensionless complexified twisted mass $m/\epsilon_2+ir/2$ so we also omit R-charges.

The 4d instanton partition function takes the form
\begin{equation}
  Z^{\text{4d}}_{\text{instanton}}(i\Sigma;q) = \sum_\lambda Z^{\text{4d}}(i\Sigma,\phi;q)
\end{equation}
where $i\Sigma$~denotes the Coulomb branch parameters and the sum ranges over a set of instanton configurations~$\phi$ labeled by some Young diagrams~$\lambda$.   We have included the instanton counting factor $q^{\lvert \lambda\rvert}$ in $Z^{\text{4d}}(i\Sigma,\phi;q)$.

We review the 2d vortex partition function in \autoref{b:2d}.  While instantons in 4d exist throughout the Coulomb branch, vortices in 2d only exist for a finite number of values $i\sigma_0$ of the Coulomb branch parameter.  These vortex partition functions are obtained (up to an overall normalization) by different choices of contours $\gamma(i\sigma_0)$ in\footnote{\label{foot:2dsimplepoles}Here we assume that the integral only has simple poles, which is the case for $U(k)$ quiver gauge theories of interest to us, with generic twisted masses.  Theories with non-trivial superpotential typically have more complicated vortex partition functions involving derivatives of Gamma functions:  these are obtained as appropriate limits of vortex partition functions without the superpotential.  In theories whose gauge group is not a product of $U(k)$ factors the vortex partition functions may be divergent series due to the lack of FI parameters.}
\begin{equation}\label{zvasintegral}
  Z^{\text{2d}}_{\text{vortex}}(i\sigma_0;z) = \int_{\gamma(i\sigma_0)} \!\!\!\!\! d \si\, Z^{\text{2d}}(i \si;z)
  = \int_{\gamma(i\sigma_0)} \!\!\!\!\! d \si\, z^{\Tr i\sigma/\epsilon_2} \frac{\prod_{w\in\text{weights}}\Gamma(-w\cdot i\sigma/\epsilon_2)}{\prod_{\alpha\in\text{roots}(G)} \Gamma(\alpha\cdot i\sigma/\epsilon_2)}
\end{equation}
where the products range over roots of the gauge group~$G$ and weights (with multiplicity) of the matter representation and there is one $z^{\Tr(i\sigma/\epsilon_2)}$ factor for each $U(1)$ gauge group.  Performing the contour integral gives, up to a different normalization,
\begin{align}
  Z^{\text{2d}}_{\text{vortex}}(i\sigma_0;z) & = \sum_{k\in\mathfrak{t}_{\text{GNO}}} Z^{\text{2d}}(i\sigma_0+k\epsilon_2;z) \\
  \label{zvmsk}
  Z^{\text{2d}}(i\sigma;z) & = \tilde{z}^{\Tr i\sigma/\epsilon_2} \prod_{w\in\text{weights}}\frac{1}{\Gamma(1+w\cdot i\sigma/\epsilon_2)} \prod_{\alpha\in\text{roots}(G)} \frac{1}{\Gamma(\alpha\cdot i\sigma/\epsilon_2)} \,.
\end{align}
The sum ranges over a lattice in the Cartan algebra~$\mathfrak{t}$ of the gauge group, defined by GNO quantization: $k$~has integer eigenvalues on any representation of~$G$ (this depends on the global structure of~$G$).  The value of~$\sigma_0$ must be chosen at the root of a Higgs branch, as explained in \autoref{b:2d}.

We abusively use the same notation $Z^{\text{2d}}(i\sigma;z)$ for quantities that differ by some exchanges $\Gamma(x)\to\Gamma(x)\frac{1}{\pi}\sin\pi x=1/\Gamma(1-x)$.  Apart from a sign that maps $z=e^{2\pi i\tau}\to\tilde{z}=e^{2\pi i\tilde{\tau}}$ with $\tilde{\tau}=\tau+\frac{1}{2}\sum_{w\in\text{weights}}w$, the extra sines take the same value for all $k\in\mathfrak{t}_{\text{GNO}}$ and only affect the normalization.  Our choice in \eqref{zvasintegral} has many infinite lattices of poles, whose sum of residues give vortex partition functions, while our choice in \eqref{zvmsk} has no pole.  Later we make intermediate choices to keep expressions short.

To generalize the 4d--2d factor in~\eqref{4d--2d-fundamental-interaction} to the case
where the 2d matter lives in an arbitrary representation of the 4d gauge group, let
us first recall the expression for the VEV of the operator $\Tr_\square e^{i\Phi}$ in
a particular instanton configuration:
\beq
\ll \Tr_\square e^{i\Phi} \rr_\la =
\sum_{A=1}^N e^{i \Sigma_A}-(1-e^{\epsilon_1})(1-e^{\epsilon_2})\sum_{I=1}^K e^{\phi_I - \epsilon_1/2-\epsilon_2/2} =
\sum_{x_l^+} e^{x_l^+} - \sum_{x_l^-} e^{x_l^-}
\eeq
where $\Sigma_A$ and $\phi_I$ correspond to some particular 4d gauge factor $U(N_i)$.
Then, using (\ref{4d2d}), the 4d--2d mixing part and the 2d perturbative contribution can be rewritten as:
\begin{multline}
\prod_{w\in\text{weights}} \prod_{l} \prod_{n=1}^{+\infty} 
\frac{x_l^+ +w\cdot i\sigma+\epsilon_2 n}{x_l^- +w\cdot i\sigma+\epsilon_2 n}\\
= \prod_{w\in\text{weights}} \biggl[ \prod_{A=1}^N\frac{1}{\Gamma(1+(i\Sigma_A+w\cdot i\sigma)/\epsilon_2)}
\prod_{I=1}^K \frac{\phi_I+w\cdot i\sigma+\frac{\epsilon_1+\epsilon_2}{2}} {\phi_I+w\cdot i\sigma+\frac{\epsilon_1+\epsilon_2}{2}-\epsilon_1} \biggr] \,.
\end{multline}

If the desired representation is defined by a Young diagram $\mu$, the
generalization of the above formula is given by the supersymmetric Schur
polynomial\footnote{We need to use supersymmetric Schur polynomials because we are
  dealing with virtual bundles.}:
\beq
\Tr_\mu e^{i \Phi} = s_\mu (e^{x_1^+},e^{x_2^+},\ldots|-e^{x_1^-},-e^{x_2^-},\ldots) =
\sum_{y_l^+} e^{y_l^+} - \sum_{y_l^-} e^{y_l^-} \,.
\eeq
Supersymmetric Schur polynomials are defined as
\beq
s_\mu (\vec{a}|\vec{b}) = \sum_{\eta}s_{\mu/\eta}(\vec{a}) s_{\eta^t}(\vec{b})
\eeq
where the sum over $\eta$ goes over Young diagrams $\eta$ inside $\mu$.  Equivalently, these polynomials are defined by requiring $s_\mu (\vec{a}|\emptyset)=s_\mu ( \vec{a})$ and the cancellation property, that is,\break $s_\mu (a_0, \vec{a}|{-a_0},\vec{b})=s_\mu (\vec{a}|\vec{b})$.

Alternatively, we can rewrite $\Tr_\mu e^{i\Phi}$ in terms of $\Tr_\square e^{k i\Phi}$ for various $k$ and compute them using (\ref{chiral}).  The two approaches are equivalent because $\Tr_\square e^{k i\Phi}$ is equal to the supersymmetric power sum
\beq
p_k=\Tr_\square e^{k i\Phi} = \sum_{x_l^+} e^{k x_l^+} - \sum_{x_l^-} e^{k x_l^-} \,,
\eeq
and the usual and supersymmetric Schur polynomials take the same form when expressed in terms of the usual and supersymmetric power sums respectively.  For example, for the symmetric representation {\scriptsize\yng(2)} both methods lead to:
\beq
  \ll \Tr_{\text{\tiny\yng(2)}} e^{i \Phi} \rr_\la = \cfrac{1}{2} \l \ll \Tr_\square e^{i \Phi} \rr_\la^2 + \ll \Tr_\square e^{2 i \Phi} \rr_\la \r
\eeq
This expression reproduces the correct contribution from a hypermultiplet in the symmetric representation \cite{mrep}.

The corresponding 4d--2d interaction and 2d perturbative contribution are given by
\beq
  \ll Z^{\text{2d}}(i\sigma;z)\rr_\la = z^{\Tr i\sigma/\epsilon_2} \prod_{\!\!\alpha\in\text{roots}(G)\!\!} \cfrac{1}{\Gamma(\alpha \cdot i\sigma/\epsilon_2)} \prod_{w\in\text{weights}} \prod_l \prod_{n=1}^{+\infty} \frac{y_l^+ +w\cdot i\sigma+\epsilon_2 n}{y_l^- +w\cdot i\sigma+\epsilon_2 n} \,.
\eeq
In the zero-instanton sector there are no $y^-$ and the infinite product reproduces the $1/\Gamma$ 2d perturbative contribution.  Contrarily to the case of a fundamental representation, splitting into a 2d perturbative contribution (zero instantons) and an interaction factor is not very instructive.

The instanton--vortex partition function is finally
\begin{equation}
  Z^{\text{4d--2d}}_{\text{instanton--vortex}}(i\Sigma,i\sigma_0;q,z) = \sum_{\lambda,k} Z^{\text{4d}}(i\Sigma,\phi;q) \ll Z^{\text{2d}}(i\sigma_0+k\epsilon_2;z)\rr_\lambda
\end{equation}
up to an overall normalization\footnote{In some cases, the sum over $\la$ includes additional configurations apart from the standard instanton Young diagrams.  See \autoref{trapped} for details.}.  As we will see in concrete examples, in the presence of instantons it is important to include negative vorticities~$k$ in the sum.

\section{Comparison with the Higgsing procedure}\label{sec:Higgs}

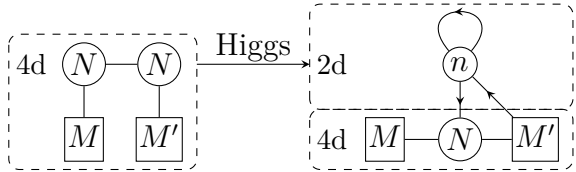
\begin{figure}
\begin{center}
\begin{tikzpicture}[inner sep=1pt,minimum size=2.5ex]
\tikzset{->-/.style 2 args={decoration={markings, mark=at position #1 with {\arrow[rotate=#2]{stealth}}},postaction={decorate}}}
\node[draw] (M1) at (0,-1) {$\strut M$};
\node[draw,circle] (N1) at (0,0) {$N$};
\node[draw,circle] (N2) at (1,0) {$N$};
\node[draw] (M2) at (1,-1) {$\strut M'$};
\draw (M1)--(N1)--(N2)--(M2);
\node at (-0.7,0) {\small 4d};
\draw[dashed,rounded corners] (-1,0.4) rectangle (1.5,-1.4);
\draw[-{stealth}] (1.5,0) to node [pos=0.5,above] {Higgs} (3,0);
\begin{scope}[shift={(4,-1)}]
\node[draw] (m1) at (0,0) {$\strut M$};
\node[draw,circle] (nc) at (1,0) {$N$};
\node[draw] (m2) at (2,0) {$\strut M'$};
\draw (m1)--(nc)--(m2);
\node at (-.7,0) {\small 4d};
\node[draw,circle] (n) at (1,1) {$n^{}$};
\node at (-.7,1) {\small 2d};
\draw[->-={.65}{0}] (n)--(nc);
\draw[->-={.65}{0}] (m2)--(n);
\draw[->-={.55}{-7}] (n) to [loop,in=130,out=50,distance=5ex] (n);
\draw[dashed,rounded corners] (-1,.4) rectangle (2.5,-.4);
\draw[dashed,rounded corners] (-1,.4) rectangle (2.5,1.8);
\end{scope}
\end{tikzpicture}
\end{center}
\caption{\label{fig:HiggsSUNSUN}We Higgs an $SU(N)\times SU(N)$ quiver gauge theory and obtain 4d $\Nsusy=2$ $SU(N)$ SQCD with $M+M'$ fundamental hypermultiplets, coupled to a 2d $\Nsusy=(2,2)$ theory with gauge group $U(n)$ and one adjoint, $M'$~fundamental and $N$~antifundamental chiral multiplets.}
\end{figure}

In this section we consider the 4d $\Nsusy=2$ $SU(N)\times SU(N)$ quiver gauge theory with one bifundamental hypermultiplet and $M,M'\leq N$ fundamental hypermultiplets for the two $SU(N)$.  We Higgs the gauge group down to the diagonal $SU(N)$ through a non-trivial vortex configuration.  Generally, this gives rise to a pair of intersecting surface defects in the $(x^1,x^2)$ and $(x^3,x^4)$ planes \cite{Gomis:2016ljm,pp}, described by two-dimensional gauge theories coupled through zero-dimensional matter at the origin.

The goal here is to confirm our prescription based on \cite{pp}  for the instanton--vortex partition function,
\begin{equation}\label{zivasvev}
  Z^{\text{4d--2d}}_{\text{instanton--vortex}} = \langle Z^{\text{2d}} \rangle_{\text{4d}} \,.
\end{equation}

\subsection{Contour prescription}

We consider $U(N)\times U(N)$ instantons\footnote{The $U(1)$~contribution can be decoupled as usual.  We choose $U(1)$ charges according to the quiver
  \begin{tikzpicture}[->-/.style 2 args={decoration={markings, mark=at position #1 with {\arrow[rotate=#2]{stealth}}},postaction={decorate}}]
    \node[draw](M1)at(0,0){$\scriptstyle M$};
    \node[draw,circle](N1)at(1,0){\clap{$\scriptstyle N$}};
    \node[draw,circle](N2)at(2,0){\clap{$\scriptstyle N$}};
    \node[draw](M2)at(3,0){$\scriptstyle M'$};
    \draw[->-={.65}{0}](M1)--(N1);
    \draw[->-={.65}{0}](N1)--(N2);
    \draw[->-={.65}{0}](N2)--(M2);
  \end{tikzpicture}:
  hypermultiplets are in $M(\square,1)+(\overline{\square},\square)+M'(1,\overline{\square})$ of $U(N)\times U(N)$.} in the Omega background, with gauge equivariant parameters $(i\Sigma,i\Sigma')$ in the Cartan algebra of $U(N)\times U(N)$.  The usual integral representation of the Nekrasov partition function reads\footnote{A mass parameter for the bifundamental hypermultiplet could be introduced by shifting $\phi'$ and $i\Sigma'$ simultaneously.  We also shifted masses $iM_f$ and $iM'_f$ to lighten formulas.}
\begin{multline}
  Z^{\text{4d}\ U(N)\times U(N)}_{\text{instanton}} = \sum_{K,K'\geq 0} \frac{q^K}{K!} \frac{q'^{K'}}{K'!} \int d^K\!\phi \, d^{K'}\!\!\phi' \Biggl[
  \\
  \prodprime_{I,J=1}^K \frac{\phi_{IJ}(\phi_{IJ}+\epsilon_1+\epsilon_2)}{(\phi_{IJ}+\epsilon_1)(\phi_{IJ}+\epsilon_2)}
  \prod_{I=1}^K \frac{\prod_{f=1}^{M}(\phi_I-iM_f) \prod_{A=1}^N(-\phi_I+i\Sigma_A'+\frac{\epsilon_1+\epsilon_2}{2})}{\prod_{A=1}^N[(\phi_I-i\Sigma_A+\frac{\epsilon_1+\epsilon_2}{2})(-\phi_I+i\Sigma_A+\frac{\epsilon_1+\epsilon_2}{2})]}
  \\
  \times \prodprime_{I,J=1}^{K'} \frac{\phi'_{IJ}(\phi'_{IJ}+\epsilon_1+\epsilon_2)}{(\phi'_{IJ}+\epsilon_1)(\phi'_{IJ}+\epsilon_2)}
  \prod_{I=1}^{K'} \frac{\prod_{f=1}^{M'}(iM'_f-\phi'_I) \prod_{A=1}^N(\phi'_I-i\Sigma_A+\frac{\epsilon_1+\epsilon_2}{2})}{\prod_{A=1}^N[(\phi'_I-i\Sigma'_A+\frac{\epsilon_1+\epsilon_2}{2})(-\phi'_I+i\Sigma'_A+\frac{\epsilon_1+\epsilon_2}{2})]}
  \\
  \times \prod_{I=1}^{K'} \prod_{J=1}^K \frac{(\phi'_I-\phi_J+\epsilon_1)(-\phi'_I+\phi_J-\epsilon_2)}{(\phi'_I-\phi_J)(-\phi'_I+\phi_J-\epsilon_1-\epsilon_2)}
  \Biggr] \,,
\end{multline}
where $\phi_{IJ}=\phi_I-\phi_J$ and $\phi'_{IJ}=\phi'_I-\phi'_J$ and $\prodprime$ means that vanishing factors $\phi_{II}$ and $\phi'_{II}$ are omitted.  We assume that masses $M_f$ and $M'_f$ are generic.  The contour is chosen to surround poles at
\begin{equation}\label{stdphiphipole}
  \begin{aligned}
    \{\phi_I\} & = \left\{ i\Sigma_A+(r-\tfrac{1}{2})\epsilon_1+(s-\tfrac{1}{2})\epsilon_2 \middle| 1\leq A\leq N, (r,s)\in \lambda_A \right\} \\
    \{\phi'_I\} & = \left\{ i\Sigma'_A+(r-\tfrac{1}{2})\epsilon_1+(s-\tfrac{1}{2})\epsilon_2 \middle| 1\leq A\leq N, (r,s)\in \lambda'_A \right\}
  \end{aligned}
\end{equation}
for any $2N$ Young diagrams $\lambda$ and~$\lambda'$.\footnote{As we prove in \autoref{a:nek} this set of poles can alternatively be described as taking the $-\sum\phi-\sum\phi'$ Jeffrey--Kirwan (JK) residue prescription.  The JK prescription includes additional poles due to bifundamentals, whose residue vanishes thanks to zeros in the numerator.}\footnote{The fact that the Nekrasov partition function is naturally expressed in terms of JK residues was proven more generally in~\cite{Martens:2006hu}.  We thank Francesco Sala for pointing this out.}

Fix~$\lambda$, namely a pole for~$\phi$.  Factors that only involve~$\phi$ and~$i\Sigma$ coincide with those for the pure 4d $\Nsusy=2$ $U(N)$ theory.  The remaining factors are
\begin{multline}\label{zinstsunsunextra}
  \prod_{A,B=1}^N \prod_{(r,s)\in \lambda_A} (i\Sigma'_B-i\Sigma_A-(r-1)\epsilon_1-(s-1)\epsilon_2)
  \\
  \times \int d^{K'}\!\!\phi'
  \prodprime_{I,J=1}^{K'} \frac{\phi'_{IJ}(\phi'_{IJ}+\epsilon_1+\epsilon_2)}{(\phi'_{IJ}+\epsilon_1)(\phi'_{IJ}+\epsilon_2)}
  \prod_{I=1}^{K'}
  \frac{\prod_{f=1}^{M'}(iM'_f-\phi'_I)}{\prod_{A=1}^N[(\phi'_I-i\Sigma'_A+\frac{\epsilon_1+\epsilon_2}{2})(-\phi'_I+i\Sigma'_A+\frac{\epsilon_1+\epsilon_2}{2})]}
  \\
  \times \prod_{I=1}^{K'} \prod_{A=1}^N
  \frac{
    \prod_{(r,s)\in\partial_+\lambda_A}
    (\phi'_I-i\Sigma_A-(r-\frac{3}{2})\epsilon_1-(s-\frac{3}{2})\epsilon_2)
  }{
    \prod_{(r,s)\in\partial_-\lambda_A}
    (\phi'_I-i\Sigma_A-(r-\frac{1}{2})\epsilon_1-(s-\frac{1}{2})\epsilon_2)
  }
\end{multline}
where the inner boundary $\partial_-\lambda_A\subset\lambda_A$ consists of every box such that removing it from~$\lambda_A$ gives a Young diagram, while the outer boundary $\partial_+ \lambda_A$ consists of positions where a box can be added to~$\lambda_A$.

The standard contour is equivalent to the $-\sum\phi'$ JK prescription, which picks poles such that every~$\phi'_J$ is $i\Sigma'_A+\frac{\epsilon_1+\epsilon_2}{2}$ for some $1\leq A\leq N$, or $\phi'_I+\epsilon_1$ or $\phi'_I+\epsilon_2$ for some $1\leq I<J$.\footnote{The JK prescription mandates to write $(-1,\ldots,-1)$ as a positive linear combination of $K'$~charge vectors among the charge vectors $(\ldots,0,\pm 1,0,\ldots)$ and $(\ldots,0,\pm 1,0,\ldots,0,\mp 1,0,\ldots)$ of denominator factors, then to compute the residue at the intersection of the corresponding hyperplanes.  For any such choice of linear combination one can define a partial order on $1\leq I\leq K'$ such that $I\prec J$ if the charge vector with $+1$ at position~$I$ and $-1$ at position~$J$ is used.  One can then permute the $\phi'_I$ so that $I\prec J\implies I<J$.}  Let us recall why $\phi'$ then takes the form~\eqref{stdphiphipole} of $N$~Young diagrams.  We check by induction on~$J$ that $\{\phi'_I \mid1\leq I\leq J\}$ takes that form too.  The key is that poles of $\prodprime_{I=1}^{J-1} \frac{\phi'_{IJ}(\phi'_{IJ}+\epsilon_1+\epsilon_2)}{(\phi'_{IJ}+\epsilon_1)(\phi'_{IJ}+\epsilon_2)}\big/\allowbreak\prod_{A=1}^N (-\phi'_J+i\Sigma'_A+\frac{\epsilon_1+\epsilon_2}{2})$ are along the outer boundary of each of the~$N$ Young diagrams: adding any of these boxes still gives Young diagrams.

We now close the $\phi'$~contours on the opposite side\footnote{\label{foot:u(1)factor}For $M'=N-1$ or $M'=N$ \eqref{zinstsunsunextra} has poles at infinity (the integrand behaves as $\phi_I'^{M'-2N+N}$ at large~$\phi_I'$) so the two contour prescriptions give different answers.  We expect them to only differ by a multiplicative factor affecting the $U(1)$ contributions.  It would be interesting to find this factor by correctly accounting for poles at infinity (see appendix~B of \cite{Gomis:2014eya} for such calculations for 2d vortices).} namely choose the $+\sum\phi'$ JK prescription.  Poles are now at
\begin{multline}\label{phiprimepole}
  \{\phi'_I\} =
  \left\{ i\Sigma'_A-(r-\tfrac{1}{2})\epsilon_1-(s-\tfrac{1}{2})\epsilon_2 \middle| 1\leq A\leq N, (r,s)\in \lambda'_A \right\}
  \\
  \cup \left\{ i\Sigma_A+(r-\tfrac{1}{2})\epsilon_1+(s-\tfrac{1}{2})\epsilon_2 \middle| 1\leq A\leq N, (r,s)\in \lambda_A\setminus \lambda''_A \right\}
\end{multline}
for $2N$ Young diagrams $\lambda'$ and $\lambda''$ with all $\lambda''_A\subseteq \lambda_A$.  The first set is analogous to what we found in the $-\sum\phi'$ JK prescription; it is based on the denominators $\phi'_J-i\Sigma'_A+\frac{\epsilon_1+\epsilon_2}{2}$.  The second one is also found by induction, starting from an empty set ($\lambda''=\lambda$).  At any stage of the induction, values of~$\phi'_J$ allowed by the denominator factor $\prod_A\prod_{(r,s)\in\partial_-\lambda''_A}(\phi'_J-i\Sigma_A-(r- \frac{1}{2})\epsilon_1-(s-\frac{1}{2})\epsilon_2)$ correspond to removing a box from~$\lambda''$.
The residue of~\eqref{zinstsunsunextra} at~\eqref{phiprimepole} is
\begin{multline}
  (\ref*{zinstsunsunextra})\bigr|_{\phi'=(\ref*{phiprimepole})} =
  \Biggl[
  \prod_{A,B=1}^N \frac{\prod_{(r,s)\in \lambda''_A} (i\Sigma'_B-i\Sigma_A-(r-1)\epsilon_1-(s-1)\epsilon_2)}{\prod_{(r,s)\in \lambda'_A}(i\Sigma'_B-i\Sigma'_A+r\epsilon_1+s\epsilon_2)}
  \\
  \times \prod_{I=1}^{K'} \biggl[ \prod_{f=1}^{M'}(iM'_f-\phi'_I) \prod_{A=1}^N \frac{\prod_{(r,s)\in\partial_-\lambda'_A}(\phi'_I-i\Sigma'_A+(r+\frac{1}{2})\epsilon_1+(s+\frac{1}{2})\epsilon_2)}{\prod_{(r,s)\in\partial_+\lambda'_A}(\phi'_I-i\Sigma'_A+(r-\frac{1}{2})\epsilon_1+(s-\frac{1}{2})\epsilon_2)}
  \\
  \times \prod_{A=1}^N \frac{\prod_{(r,s)\in\partial_+\lambda''_A}(\phi'_I-i\Sigma_A-(r-\frac{3}{2})\epsilon_1-(s-\frac{3}{2})\epsilon_2)}{\prod_{(r,s)\in\partial_-\lambda''_A}(\phi'_I-i\Sigma_A-(r-\frac{1}{2})\epsilon_1-(s-\frac{1}{2})\epsilon_2)} \biggr] \Biggr]_{\phi'=(\ref*{phiprimepole})}
\end{multline}
where we left some~$\phi'$ in the right-hand side to keep expressions short.

\subsection{Higgsing}

We can now Higgs: $i\Sigma'_A\to i\Sigma_A+n_A\epsilon_1$.  The denominator remains nonzero in this limit\footnote{On the contrary, in the limit $i\Sigma'_A\to i\Sigma_A+n_A\epsilon_1+m_A \epsilon_2$, with $m \neq 0$ and $n \neq 0$, the denominator can blow up.  This limit produces intersecting defects.}, while the numerator vanishes unless all $\widetilde{\lambda}'_{A1}+\widetilde{\lambda}''_{A1} \leq n_A$.  The data of $\lambda',\lambda''$ can be recast as $n=\sum_A n_A$ integers
\begin{equation}
  k_{Ar} = \begin{cases}
    - \lambda''_{Ar} & \text{for $1\leq r\leq\widetilde{\lambda}''_{A1}$,} \\
    0 & \text{for $\widetilde{\lambda}''_{A1}<r\leq n_A-\widetilde{\lambda}'_{A1}$,} \\
    \lambda'_{A(n_A+1-r)} & \text{for $n_A-\widetilde{\lambda}'_{A1}<r\leq n_A$}
  \end{cases}
\end{equation}
which sum to $K'-K$, are bounded below as $k_{Ar}\geq -\lambda_{Ar}$, and obey $k_{Ar}\leq k_{A(r+1)}$.

The (Higgsed) instanton partition function is then
\begin{equation}\label{4dhiggsed1}\compressthisequation
  Z^{\text{4d},\text{Higgsed}}_{\text{instanton}}(i\Sigma;q,q')
  = \frac{1}{Z_0} \sum_{\lambda,k}
  \Bigl[ Z^{\text{4d}}(i\Sigma,\phi; qq') Z^{\text{4d--2d}}(\phi,i\sigma) Z^{\text{2d}}(i\Sigma,i\sigma;\epsilon_2^{N-M'}/q') \Bigr]_{\substack{\phi=\phi(i\Sigma,\lambda)\\i\sigma=i\sigma(i\Sigma,k)}}
\end{equation}
where the normalization $Z_0=Z^{\text{2d}}(i\Sigma,i\sigma(i\Sigma,k=0);z)$ sets the leading term to~$1$ and
\begin{gather}
  \begin{aligned}
    \phi(i\Sigma,\lambda) & = \{i\Sigma_A+(r-\tfrac{1}{2})\epsilon_1+(s-\tfrac{1}{2})\epsilon_2 \mid 1\leq A\leq N,(r,s)\in \lambda_A\} \,,\\
    i\sigma(i\Sigma,k) & = \{i\Sigma_A+(r-1)\epsilon_1-k_{Ar}\epsilon_2 \mid 1\leq A\leq N,1\leq r\leq n_A\} \,,
  \end{aligned}
  \\\label{instantonvortexpieces}
  \compressthisequation
  \begin{aligned}
    Z^{\text{4d}}(i\Sigma,\phi;x) & = x^K \prodprime_{I,J=1}^K \frac{\phi_{IJ}(\phi_{IJ}+\epsilon_1+\epsilon_2)}{(\phi_{IJ}+\epsilon_1)(\phi_{IJ}+\epsilon_2)} \prod_{I=1}^K \frac{\prod_{f=1}^M(\phi_I-iM_f)\prod_{f=1}^{M'}(iM'_f-\phi_I)}{\prod_{A=1}^N\prod_{\pm}(\pm(\phi_I-i\Sigma_A)+\frac{\epsilon_1+\epsilon_2}{2})} \,, \mspace{-50mu}\\
    Z^{\text{4d--2d}}(\phi,i\sigma) & = \prod_{I=1}^K \prod_{j=1}^n \frac{\phi_I-i\sigma_j+\frac{\epsilon_1+\epsilon_2}{2}}{\phi_I-i\sigma_j+\frac{\epsilon_1+\epsilon_2}{2}-\epsilon_1} \,,\\
    Z^{\text{2d}}(i\Sigma,i\sigma;z) &
    = \tilde{z}^{\Tr i\sigma/\epsilon_2} \prod_{i,j=1}^n \frac{\Gamma(1+(i\sigma_j-i\sigma_i)/\epsilon_2)}{\Gamma(1+(i\sigma_j-i\sigma_i+\epsilon_1)/\epsilon_2)} \prod_{j=1}^n \frac{\prod_{f=1}^{M'}\Gamma((iM'_f+\frac{\epsilon_2-\epsilon_1}{2}-i\sigma_j)/\epsilon_2)}{\prod_{A=1}^N\Gamma(1+(i\Sigma_A-i\sigma_j)/\epsilon_2)}
    \mspace{-50mu}
  \end{aligned}
\end{gather}
with $\tilde{z}=(-1)^Nz$.
While $Z^{\text{2d}}$~vanishes when any~$k_{Ar}$ is negative, this is partly cancelled by~$Z^{\text{4d--2d}}$: it combines with the last factor of~$Z^{\text{2d}}$ into the (regularized) infinite product
\begin{multline}\label{z4d2dcombined}
  Z^{\text{4d--2d}} \prod_{A=1}^N \prod_{j=1}^n \frac{1}{\Gamma(1+(i\Sigma_A-i\sigma_j)/\epsilon_2)}
  = \prod_{j=1}^n \prod_{A=1}^N \frac{\prod_{s=1}^{\lambda_{A1}}((i\Sigma_A-i\sigma_j+\widetilde{\lambda}_{As}\epsilon_1+s\epsilon_2)/\epsilon_2)} {\Gamma(1+(i\Sigma_A-i\sigma_j+\lambda_{A1}\epsilon_2)/\epsilon_2)} \\
  = \prod_{j=1}^n \prod_{A=1}^N \prod_{s\geq 1}^{\text{reg.}} \Bigl(\frac{i\Sigma_A-i\sigma_j+\widetilde{\lambda}_{As}\epsilon_1+s\epsilon_2}{\epsilon_2}\Bigr) \,.
\end{multline}
Let us show that replacing the sum by a sum over all $k\in\mathbb{Z}^n$ does not add terms.  We note first that $\prod_{i,j=1}^n 1/ \Gamma(1+(i\sigma_j-i\sigma_i+\epsilon_1)/\epsilon_2)$ vanishes unless all $k_{Ar}\geq k_{A(r-1)}$, namely the sum automatically restricts to nondecreasing sequences.  Then we note that~\eqref{z4d2dcombined} vanishes unless $k_{A(\widetilde{\lambda}_{As}+1)}>-s$ for all $s\geq1$.  Since the~$k$ are nondecreasing, we deduce $k_{Ar}>-s$ whenever $r>\widetilde{\lambda}_{As}$, namely whenever $s>\lambda_{Ar}$.  This is equivalent to $k_{Ar}\geq -\lambda_{Ar}$ and we are back to the original sum.

As we now explain, this expression is in fact the instanton--vortex partition function of the 4d--2d coupled system depicted in \autoref{fig:HiggsSUNSUN}.  The 4d $\Nsusy=2$ theory is a $U(N)$ vector multiplet: $Z^{\text{4d}}$ is the usual summand of the $U(N)$ Nekrasov partition function with $M$~fundamental and $M'$~antifundamental hypermultiplets.  The 2d $\Nsusy=(2,2)$ $U(n)$ gauge theory has one adjoint, $M'$~fundamental and $N$~antifundamental chiral multiplets.  Indeed, $Z^{\text{2d}}$ is the vortex partition function of the 2d theory in the absence of 4d instantons: ignoring $k$-independent normalizations as done below~\eqref{zvmsk},

\begin{itemize}
\item $\prod_{i,j=1}^n \Gamma(1+(i\sigma_j-i\sigma_i)/\epsilon_2) \simeq(-1)^{(n-1)\lvert k\rvert}\prod_{i<j}^n ((i\sigma_j-i \sigma_i)/\epsilon_2)$ is the vector multiplet contribution,
\item $\prod_{i,j=1}^n 1/\Gamma(1+(i\sigma_j-i\sigma_i+\epsilon_1)/ \epsilon_2)$ is the contribution of an adjoint chiral multiplet of (vector) R-charge $-2\operatorname{Re}(\epsilon_1/\epsilon_2)$ and twisted mass $\operatorname{Im}(\epsilon_1/\epsilon_2)$,
\item $\prod_{A=1}^N \prod_{j=1}^n 1/\Gamma(1+(i\Sigma_A-i\sigma_j)/ \epsilon_2)$ is the contribution of $N$~antifundamental chiral multiplets of twisted masses $\Sigma_A/\epsilon_2$,
\item $\prod_{A=1}^N\prod_{f=1}^{M'}\Gamma((iM'_f- \frac{\epsilon_1}{2}+\frac{\epsilon_2}{2}-i\sigma_j)/\epsilon_2)$ is the contribution of $M'$~fundamental chiral multiplets of (vector) R-charge $1-\operatorname{Re}(\epsilon_1/\epsilon_2)$ and twisted masses $\frac{1}{2}\operatorname{Im}(\epsilon_1/\epsilon_2)-M'_f$.
\end{itemize}
Twisted masses and R-charges are fixed as follows.  The fundamental and antifundamental 2d chiral multiplets are coupled to the 2d restriction of the antifundamental 4d hypermultiplets through a cubic superpotential.  This identifies their $S[U(M')\times U(N)]$ flavor symmetry to symmetries of the 4d hypermultiplets and in particular requires gauging the $U(N)$ flavor symmetry using the 2d restriction of the 4d vector multiplet.  The R-charge $1-\operatorname{Re}(\epsilon_1/\epsilon_2)$ is due to a difference in the R-symmetry in 4d and 2d, as explained in \cite{Gomis:2016ljm}.

Finally, $Z^{\text{4d--2d}}$ expresses how the instanton background affects the contribution of antifundamental chiral multiplets to the vortex partition function\footnote{\label{trapped}For chiral multiplets in other representations of the 4d gauge group, the JK prescription for $\phi_I$ also picks poles from $Z^\text{4d--2d}$.  These contributions correspond to the trapped instantons we mentioned earlier.}.

For the matching between the Higgsing calculation and the instanton--vortex partition function it is crucial that our prescription includes all (integer) vorticities~$k$ such that the summand in~\eqref{4dhiggsed1} is nonzero.

\subsection{Higgsing for quiver gauge theories}

Now it is easy to generalize our Higgsing procedure for linear quiver gauge theories with $V$ nodes and with $U(N)$ gauge group on each node.  For definiteness we Higgs all the way down to a single 4d $U(N)$ gauge group: we Higgs $i \Sigma^{(i)}_A \to i \Sigma^{(i-1)}_A + n^{(i)}_A \epsilon_1$ for $1\leq i\leq(V-1)$ with $\Sigma^{(0)}_A= \Sigma_A$.

Similarly to the case that was discussed above, we choose the JK prescription $-\sum\phi_I$ on the first node and $+\sum\phi^{(i)}$ on all other nodes $1\le i \le(V-1)$.  As a result the poles are classified by pairs of partitions $(\lambda'^{(i)}_A,\lambda''^{(i)}_A)$ and a partition $\lambda_A$ with
\begin{multline}
  \{\phi^{(i)}_I\} =
  \left\{ i\Sigma^{(i)}_A-(r-\tfrac{1}{2})\epsilon_1-(s-\tfrac{1}{2})\epsilon_2 \middle| 1\leq A\leq N, (r,s)\in \lambda'^{(i)}_A \right\}
  \\
  \cup \left\{ i\Sigma_A+(r-\tfrac{1}{2})\epsilon_1+(s-\tfrac{1}{2})\epsilon_2 \middle| 1\leq A\leq N, (r,s)\in \lambda_A\setminus \lambda''^{(i)}_A \right\}
\end{multline}
and with the same $\phi_I$ as in (\ref{stdphiphipole}).

As in the case of two nodes the non-vanishing contribution only comes from $\widetilde{\lambda}'^{(i)}_{A1}+\widetilde{\lambda}''^{(i)}_{A1}\leq m^{(i)}_A$ where $m^{(i)}_A = \sum_{j=1}^i n^{(i)}_A$, and it is useful to recast $\lambda'$ and $\lambda''$ as
\begin{equation}
  k^{(i)}_{Ar} = \begin{cases}
    - \lambda''^{(i)}_{Ar} & \text{for $1\leq r\leq\widetilde{\lambda}''^{(i)}_{A1}$,} \\
    0 & \text{for $\widetilde{\lambda}''^{(i)}_{A1}<r\leq m^{(i)}_A-\widetilde{\lambda}'^{(i)}_{A1}$,} \\
    \lambda'^{(i)}_{A(m^{(i)}_A+1-r)} & \text{for $m^{(i)}_A-\widetilde{\lambda}'^{(i)}_{A1}<r\leq m^{(i)}_A$.}
  \end{cases}
\end{equation}
In addition to that there is another condition $k^{(i)}_{A (m^{(i)}_A+1-r)} \ge k^{(i+1)}_{A (m^{(i+1)}_A+1-r)}$ to get a non-vanishing contribution.

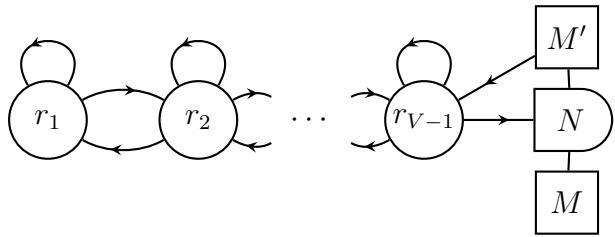
\begin{figure}
\centering
\begin{tikzpicture}[and/.style={and gate US,thick, draw,
		xshift=-.5mm}, thick,
  ->-/.style 2 args={decoration={markings, mark=at position #1 with {\arrow[rotate=#2]{stealth}}},postaction={decorate}}]
  \node[circle,draw,minimum size=2.5em] (1) at (0,0) {$r_1$};
  \node[circle,draw,minimum size=2.5em] (2) at (2,0) {$r_2$};
  \node[minimum size=2.5em] (dots) at (3.5,0) {$\cdots$};
  \node[circle,draw,minimum size=2.5em] (M1) at (5,0) {$\mathclap{r_{V-1}}$};  
  \node [and,minimum size=2.0em] (N) at (7,0)  {$N$};
  \node [draw,minimum size=2.0em] (Mp) at (6.9,1.1)  {$M'$};
  \node [draw,minimum size=2.0em] (Mf) at (6.9,-1.1)  {$M$};
  \draw[->-={.65}{5}] (1) to [loop,in=150,out=30,distance=2ex] (2);
  \draw[->-={.65}{5}] (2) to [loop,in=330,out=210,distance=2ex] (1);
  \draw[->-={.65}{5}] (2) to [loop,in=150,out=30,distance=1ex] (dots);
  \draw[->-={.65}{5}] (dots) to [loop,in=330,out=210,distance=1ex] (2);
   \draw[->-={.65}{5}] (dots) to [loop,in=150,out=30,distance=1ex] (M1);
  \draw[->-={.65}{5}] (M1) to [loop,in=330,out=210,distance=1ex] (dots);
  \draw[->-={.65}{0}] (M1) -- (N);
  \draw[->-={.65}{0}] (Mp) -- (M1);
    \draw[->-={2.65}{0}] (Mp) -- (N);
   \draw[->-={2.65}{0}] (Mf) -- (N);
  \draw[->-={.65}{-10}] (1) to [loop,in=120,out=60,distance=5ex]   (1);
   \draw[->-={.65}{-10}] (2) to [loop,in=120,out=60,distance=5ex]   (2);
    \draw[->-={.65}{-10}] (M1) to [loop,in=120,out=60,distance=5ex]   (M1);
\end{tikzpicture}
\caption{\label{4d2ddefect_quiver}Quiver diagram of the 4d--2d coupled system which appears after the Higgsing procedure for a linear quiver 4d gauge theory.}
\end{figure}

Finally, the Higgsed partition function has the following form
\begin{equation}
Z^{\text{4d},\text{Higgsed}}_{\text{instanton}}(i\Sigma;q^{(i)})
= \frac{1}{Z_0} \sum_{\lambda} \sum_{\vec{k}} Z^{\text{4d}}(i\Sigma,\phi
;x) Z^{\text{4d--2d}}(\phi,i\sigma^{(V-1)}) Z^{\text{2d}}(i\Sigma,i
\vec{\sigma};\vec{z})
\end{equation}
where $x=\prod_{i=0}^{V-1}q^{(i)}$ and $z_{V-1}=\epsilon_2^{N-M'}/q^{(V-1)}$ and other $z_i=1/q^{(i)}$, the contributions $Z^{ \text{4d}}(i\Sigma,\phi;x)$ and $Z^{\text{4d--2d}}(\phi,i\sigma^{(V-1)})$ are the same as above, while
\begin{gather}
  \begin{aligned}
    \phi = \phi(i\Sigma,\lambda) & = \{i\Sigma_A+(r-\tfrac{1}{2})\epsilon_1+(s-\tfrac{1}{2})\epsilon_2 \mid 1\leq A\leq N,(r,s)\in \lambda_A\} \,,\\
    i\sigma^{(i)} = i\sigma^{(i)}(i\Sigma,k) & = \{i\Sigma^{(i)}_A+(r-1)\epsilon_1-k^{(i)}_{Ar}\epsilon_2 \mid 1\leq A\leq N,1\leq r\leq m^{(i)}_A\} \,,
  \end{aligned}
  \\
  \!\begin{multlined}[t]
  Z^{\text{2d}}(i\Sigma,i\vec{\sigma};\vec{z}) =
  \Bigl(\tilde{z}_{V-1}^{\Tr i\sigma^{(V-1)}} \prod_{l=1}^{V-2} \tilde{z}_l^{\Tr i\sigma^{(l)}}\Bigr)
  \prod_{j=1}^{r_{V-1}} \frac{\prod_{f=1}^{M'}\Gamma((iM'_f-\frac{\epsilon_1}{2}+\frac{\epsilon_2}{2}-i\sigma^{(V-1)}_j)/\epsilon_2)}{\prod_{A=1}^N\Gamma(1+(i\Sigma_A-i\sigma^{(V-1)}_j)/\epsilon_2)} \\
    \times\prod_{l=1}^{V-1} \prod_{i,j=1}^{r_{l}} \frac{\Gamma(1+(i\sigma^{(l)}_j-i\sigma^{(l)}_i)/\epsilon_2)}{\Gamma(1+(i\sigma^{(l)}_j-i\sigma^{(l)}_i+\epsilon_1)/\epsilon_2)} 
    \prod_{l=1}^{V-2} \prod_{i=1}^{r_{l}} \prod_{j=1}^{r_{l+1}} \frac{\Gamma(1+(i\sigma^{(l+1)}_j-i\sigma^{(l)}_i+\epsilon_1)/\epsilon_2)}{\Gamma(1+(i\sigma^{(l+1)}_j-i\sigma^{(l)}_i)/\epsilon_2)}
  \end{multlined}
\end{gather}
with $r_l = \sum_A m^{(l)}_A$ and $\tilde{z}_{V-1}=(-1)^N z_{V-1}$ and $\tilde{z}_l=(-1)^{r_{l+1}+r_{l-1}}z_l$.  The resulting
partition function coincides with the partition function of the coupled
4d--2d system in \autoref{4d2ddefect_quiver}.  It is precisely the theory
that lives on non-Abelian strings which appear after Higgsing.  The
integers $k^{(i)}_A$ parametrize vortex numbers of the $i$-th node
and again can be negative.  The first ratio of Gamma functions in
$Z^{\text{2d}}$ is due to the interaction with the bulk while the
second and the third one are due to $\Nsusy=(4,4)$ vector
multiplets and bifundamental hypermultiplets deformed by the presence
of the $\Omega$-background $\epsilon_1$ in the orthogonal direction.

It is also possible to flip the contours on different nodes.  Such partition function must be related to defects coupled to 4d quiver gauge theories.  In particular if we choose $-\sum\phi_I$ for the first $P$ nodes and $+\sum\phi_I$ for the other ones we obtain a 4d quiver gauge theory with $P$ nodes with a defect that can be described as a 2d quiver theory with $(V-P)$ nodes coupled to a node of the bulk.

\subsection{Coupling 3d to 5d and U(1) factors}

Let us comment on how our prescription can be generalized for 5d--3d systems.  It is well known that in order to obtain the partition function of a five-dimensional gauge theory on $\RR_{\Omega}^4 \times S^1_\beta$ with a codimension-two surface operator we simply have to use K-theoretic formulas instead of cohomological ones \cite{nekrasov,dgh}, i.e., we should replace all rational multipliers in the integral representation of the Nekrasov partition function (or in the contribution of a given fixed point) by their trigonometric analogues:
\begin{equation}
  x_i \to 2 \sinh{\biggl(\frac{\beta x_i}{2}\biggr)}
\end{equation}
where $\beta$ is the deformation parameter which coincides with the $S^1_\beta$ circumference.  Our previous analysis can be also applied in this case except the subtleties associated with flipping the contour.  For 4d--2d systems with some matter content it may happen that there are poles at infinity and two contour prescriptions give different answers as was mentioned in \autoref{foot:u(1)factor}.  For 5d--3d systems there are always poles at infinities even without matter.  However we expect that the two answers in fact differ by some simple $U(1)$ factors both for 4d--2d and 5d--3d systems.

To shed some light on the appearance of these $U(1)$ factors let us analyze the simplest example of a 4d/5d $U(1)$ gauge theory coupled to a 2d/3d $U(1)$ theory on the defect.  Comparing two different answers for two choices of the contour for the 4d--2d system we obtain the following relation
\begin{equation}
Z^{\text{4d--2d}}_{\text{instanton--vortex}} (\Lambda, z) = \exp
\biggl(\frac{\epsilon_1+\epsilon_2}{\epsilon_1 \epsilon_2} z
\biggr) Z^{\text{4d,Higgsed}}_{\text{instanton}}(\Lambda, z)
\end{equation}
where $\Lambda$ and $z$ are the instanton parameter and vortex parameter respectively.

For the 5d--3d system the prefactor is a little bit more complicated.  We claim that it has the following form
\begin{equation}
Z^{\text{5d--3d}}_{\text{instanton--vortex}} (\Lambda, z) = \exp
\biggl(\sum_{n=1}^\infty \frac{(q_1^2 q_2)^n-q_1^n}{n(1-q_1^n)(1-q_2^n)} z^n \biggr) Z^{\text{5d,Higgsed}}_{
\text{instanton}}(\Lambda, z)
\label{eq:u(1)factor}
\end{equation}
where $q_{1,2} = e^{\beta \epsilon_{1,2}}$.

After it is written in this form it becomes clear what is the interpretation of this $U(1)$ factor.  The whole 5d--3d partition function is the generating function of open/closed BPS states of our system (see \cite{refvertex,dgh}). The contribution of the ensemble of states that comes from each closed BPS single-particle state with weight $(j_L, j_R)$ under $SU(2)_L \times SU(2)_R$ has the form
\begin{equation}
\exp\biggl( \sum_{n=1}^\infty \frac{(-1)^{2 j_L+2 j_R} q_1^{n(j_L+j_R)}q_2^{n(j_L-j_R)} Q^n}{n(1-q_1^n)(1-q_2^n)}\biggr)
\end{equation}
with an appropriately chosen mass parameter $Q$.  Our $U(1)$ factor has precisely this form, which tells us that the two systems in (\ref{eq:u(1)factor}) differ by some simple closed BPS states in the same way as the $U(N)$ Nekrasov partition function differs from the $SU(N)$ partition function.  We expect that a similar relation holds for more complicated examples.

\subsection{Instanton--negative-vortices}
\label{sec:negativebranes}

It is instructive to compare our setting with the non-perturbative
effects in $SU(N)$ $\Nsusy=1$ SYM theory on $\RR^3
\times S^1$ which we shall briefly review now \cite{davies}.  The
$\Nsusy=1$ SYM theory is the worldvolume theory on $N$ D4 branes
extended along $(x_0,x_1,x_2,x_3,x_4)$ and stretched between
an NS5 brane and an NS5$'$ brane (these have different directions), and
this brane setup can be wrapped around $S^1$.  The instantons are
represented by Euclidean D0 branes stretching between NS5 and NS5$'$
along $x_4$ and localized on $S^1$.  To recognize their
compositeness, T-dualize along the circle $S^1$ of radius~$\beta$,
whose coordinate we will denote~$x_0$.

The T-dual of the brane configuration without instantons consists of
$N$ D3 branes stretching between NS5 and NS5$'$ and localized at points
along the dual of~$x_0$.  An instanton becomes a Euclidean D1 brane
wrapping the circle; it gets fractionalized into $N$ parts stretching
between two neighbor D3 branes.  Semiclassically there is a moduli space
parametrized by the VEV $\ll A_0 \rr$ or by the
holonomy
\beq
W(A_0)= \diag(\exp(ia_1),\dots, \exp(ia_N)) \,.
\eeq
It is these eigenvalues of holonomy which fix the positions of the D3
branes along the dual circle.  The SYM theory on $\RR^3\times
S^1$ develops nonperturbatively a superpotential for the complexified
Wilson loop along $S^1$ which imposes a nonvanishing VEV for the
holonomy hence breaks $SU(N)$ to $U(1)^{N-1}$.  The quantum theory has
$N$ vacua, which differ by the value of the $\theta$~angle.  There can
be BPS domain walls separating vacua, with a nontrivial profile of the
axion.

To identify the constituents it is useful to note that the brane
configuration is of the ``Nahm type'' \cite{diaconescu} involving a
$p$-brane transverse to $(p+2)$-branes, that is, we can expect
monopole-type solutions to the equations of motion on the D3
worldvolume.  From the D3 worldvolume point of view, the ends of D1
branes are 3d monopoles.  These are so-called instanton--monopole
solutions to the equations of motion with two quantum numbers
$(Q_{\text{inst}},Q_{\text{mon}})$ corresponding to the monopole and
instanton charges
\beq
Q_{\text{inst}}=\int dx_0 \ d^3x \Tr F\tilde{F}
\quad\text{and}\quad Q_{\text{mon}}=\int d^3x \, \epsilon_{ijk}
\partial_i F_{jk} \,.
\eeq
The charges of the $i$-th instanton--monopole are
\beq
Q_{\text{inst}} = \frac{1}{N} \quad\text{and}\quad Q_{\text{mon}}=
\alpha_i
\eeq
where $\alpha_i$, $1\leq i<N$, are the simple roots of $SU(N)$.  The
explicit solutions are independent of $x_0$ and for $SU(2)$ the
solution looks as follows ($\sigma^a$~are Pauli matrices):
\begin{gather}
\label{bps1}
A_0(x)= \left( v|x|\coth(v|x|-1) \right)
\frac{x_a\sigma^a}{2i|x|^2} \,,
\\
\label{bps2}
A_\mu = \epsilon_{\mu\nu a} \left( 1 - \frac{v|x|}{\sinh(v|x|)}
\right) \frac{x^{\nu}\sigma^a}{2i|x|^2} \,.
\end{gather}
The real parameter $v$ is a coordinate on the classical moduli space.

There is also one more solution to the equation of motion: the so-called
``instanton--KK-monopole'' with quantum numbers
\beq
Q_{\text{inst}}=\frac{1}{N} \quad\text{and}\quad Q_{\text{mon}}= -
\sum_i \alpha_i \,.
\eeq
It is possible to obtain the corresponding solutions to the equations
of motion for $SU(2)$ from the solutions \eqref{bps1}--\eqref{bps2} as
follows: shift the coordinate on the moduli space $v\to\frac{2\pi}{
\beta}-v$ then perform the gauge transformation
\beq
U_s=\exp \left( \frac{\pi x_4 \tau^3}{i\beta} \right) \,.
\eeq
The instanton can be considered as a composite object built from the
$(N-1)$ instanton--monopoles and one instanton--KK-monopole.  It has
vanishing total monopole charge and one unit of instanton charge.

The instanton--monopoles play a prominent role in the nonperturbative physics of
$\Nsusy=1$ SYM on $\RR^3\times S^1$.  Their key effect is the generation of a
non-perturbative superpotential in the theory, which lifts the moduli space and
results in a number of ground states consistent with the Witten index.  These
configurations have two fermionic zero modes, $N$~times fewer than for
instantons.  This number fits perfectly with the zero mode counting for the
gluino condensate and nowadays it is widely accepted that these
instanton--monopoles or fractional instantons are responsible for the formation
of the gluino condensate.

Let us turn to our brane configurations for 4d theories.  To make the comparison
closer consider~$N$ D2 branes attached to D4 branes and localized at some points
$x_{4,i}$, $i=1\dots N$ along the $x_4$~coordinate.  The instanton is
represented by a Euclidean D0 brane extended along~$x_4$.  It is clear that the
geometry is Nahm-like once again since we have D0 brane transverse to D2 branes,
which can be split into fragments.  In the previous case the direction along
which the $N$ D2 branes are localized is compact.  In the present case we could
make the coordinate $x_4$ compact by considering the elliptic $\Nsusy=2^*$
model or another $\widehat{A}$ affine quiver; then $x_4$ and the M-theory
coordinate form the M-theory torus whose parameter is identified with the
complexified coupling constant.

In spite of the geometrical similarity the interpretation is essentially
different since the compact coordinate is part of space-time in the former case
while it is part of the internal manifold in the latter.  In principle we could
perform a T-duality transform in the $x_4$ direction but there is no need for
this step.  We are interested in our paper in the theory on the defect and
therefore we have discussed in the body of the text the effects of the classical
solutions to the equations of motion in the D2 worldvolume theory which
correspond to the ends of the D0 branes.

As above it is evident that we can split the D0 brane into fragments stretched
between D2 branes.  Each fragment has two quantum numbers as before but their
interpretation is different.  The instanton quantum number
$Q_{\text{inst}} = \int dx_7dx_4dx_1dx_2 TrF\tilde{F}$ can be seen upon the
T-duality in $x_4$ direction.  The second quantum number is the monopole charge
in the D2 worldvolume theory.  The solution in the 3d theory becomes an
instanton in the 2d sigma model, which we refer to as vortex in the text.  From
the 4d viewpoint an instanton can be trapped by the surface operator and get
fractionalized.  One of the constituents has a negative ``monopole'' quantum
number (for us, vorticity) to make the total charge vanish.  This fragment can
be called an ``instanton--negative-vortex'' (see \autoref{d4_def}) and it is an
analogue of the instanton--KK-monopole in SYM on $\RR^3\times S^1$.  Let us
emphasize that all constituents sit at one point in $\RR^4$ on the surface
defect.

To pursue the analogy further we could comment on the field theory realizations
of our constituents. A simplification occurs when the D2 brane is almost
infinite. Physically it means that we consider the Gukov--Witten defect or
equivalently a non-Abelian string with an infinite tension and consider the
limit when the instanton is completely trapped by the defect. All fields in the
instanton solution are then localized inside the defect and the situation is
analogous to a monopole trapped by a non-Abelian string \cite{tong}. In this
limit, an approximate solution of the equations of motion for the $SU(2)$ case
is \eqref{bps1} for the field $A_4$ and \eqref{bps2} for the gauge field
$A_\mu$, where $\mu$ runs over the D2 worldvolume coordinates
$(x_1,x_2,x_7)$. The solution does not depend on $x_4$ coordinate. If the D2
brane has a finite extent in $x_7$ the symmetric solution cannot be applied and
has to be modified.

Finally we could question how the positions of D2 branes along $x_4$ are
fixed. In $\Nsusy=1$ SYM on $\RR^3\times S^1$ this fixing takes place due to a
nonperturbative superpotential for the complexified Wilson loops, induced by the
instanton--monopole constituents~\cite{davies}. The positions of D2 branes
correspond to a minimum of the superpotential. In our case we can suggest that a
similar role is played by the Hamiltonian of the holomorphic integrable systems
governing the low-energy effective action. Indeed the coordinates of D2 branes
along $x_4,x_{11}$ (more accurately on the spectral curve) play the role of the
complex coordinates in the Hamiltonian systems (or rather, the angle-like
variables). The VEV of scalars parametrize the base of the Lagrangian fibration
(see \cite{gm} for a review). Contrary to the $\RR^3\times S^1$ case here in the
theory without $\Omega$-deformation we should not fix the positions of D2 branes
but the corresponding trajectories of the holomorphic Hamiltonian systems at
fixed values of the integrals of motion. It would be interesting to relate this
Hamiltonian with the nonperturbative superpotential for the complexified
holonomy of $A_6$ and the condensate formation on the defect in the mirror
description.

To complete this brief discussion of the analogy with $\Nsusy=1$ SYM theory on
$\RR^3\times S^1$ a few final remarks are in order. First note that we
considered above the case when the $\Omega$-deformation is switched off and it
would be interesting to describe the solution to the equation of motion
corresponding to an instanton--monopole and instanton-KK-monopole in the
deformed case.  Secondly, it would be interesting to perform a detailed
comparison with the ADHM-like description of instantons in the Higgs phase
\cite{nitta}.

And finally remark that our solution for the $A_4$ field involves a nontrivial
profile in the space-time coordinates $(x_1,x_2)$ on the defect worksheet. It is
this field which is important for the fractionalization of the instanton on the
defect. In the 2D $\mathbb{C}P^N$ model which can be considered as the
worldsheet theory on the defect when the holonomy of the $A_4$ is switched off
the fractionalization of the instanton into the kink-instanton and
KK-kink-instantons occurs only in $\RR^1\times S^1$ geometry (see \cite{dunne}
for the discussion on this issue). In our case there is no compact coordinate in
2D defect theory but there is the additional field which does the job. Note the
holonomy of $A_4$ is closely related to the $\theta$-term or axion in more
general situations. Hence presumably our solution tells that we have a
non-trivial axion profile on the defect induced by the vortex. This
interpretation deserves a further clarification.

\section{Comparison with orbifold calculations}
\label{sec:orbifold}

In this section we test our prescription for the Gukov--Witten (GW)
surface operator breaking the $SU(N)$ 4d gauge group to $S[U(n_1)
\times \dots \times U(n_M)]$. It is defined by imposing a singularity
\begin{equation}
\label{GWmonodromy}
A_\mu dx^{\mu}\sim\diag(\underbrace{\alpha_{(1)},
\ldots,\alpha_{(1)}}_{\text{$n_1$ times}}, \underbrace{\alpha_{(2)},
\ldots,\alpha_{(2)}}_{\text{$n_2$ times}}, \ldots, \underbrace{
\alpha_{(M)},\ldots,\alpha_{(M)}}_{\text{$n_M$ times}}) id\theta
\end{equation}
for the gauge field near the surface operator, and we set $0\leq
\alpha_{(M)}<\cdots<\alpha_{(1)}<2\pi$ using a gauge transformation.
As far as its instanton partition function is concerned, the 4d
$\Nsusy=2$ $SU(N)$ theory with this GW surface operator is
equivalent to the same theory on an orbifold $\mathbb{C}\times(
\mathbb{C}/\mathbb{Z}_M)$ where $\mathbb{Z}_M$ also acts as the
gauge transformation $(\underbrace{\omega,\ldots}_{n_1\text{ times}},\underbrace{\omega^2,\ldots}_{
n_2\text{ times}},\ldots,\underbrace{\omega^M,\ldots}_{n_M\text{ times}})\in U(N)$ with $\omega=e^{2\pi i/M}$ (\mbox{see
\cite{orbi}}).

\subsection{Review of the orbifold construction}

Let us briefly recall how Kanno and Tachikawa derived the instanton partition function for this orbifold in \cite{orbi}.

First we review the case with no orbifold. The standard $d$-instanton
moduli space of $U(N)$~instantons is $\mathcal{M}_{\text{ADHM}}=\{(A,B,P,Q)
\mid[A,B]+PQ=0\}/\!/GL(d,\mathbb{C})$ where $A,B:V\to V$, $P:W\to V$,
$Q:V\to W$ and $\dim_{\mathbb{C}}V=d$ and $\dim_{\mathbb{C}}W=N$. Fixed
points under the global symmetries $(A,B,P,Q)\to(e^{\epsilon_1}A,e^{\epsilon_2}B,P,e^{\epsilon_1+\epsilon_2}Q)$ and $(A,B,P,Q)
\to(A,B,Pg^{-1},gQ)$ (for $g\in U(N)$) are labeled by
$N$-tuples of Young diagrams~$\lambda$. Denoting by $i\Sigma$ the
$U(N)$ equivariant parameters, the characters of $V$ and $W$ (abusively
denoted by $V$ and~$W$ too) are
\begin{equation}
V = \sum_{A=1}^N \sum_{(r,s)\in\lambda_A} e^{i\Sigma_A+(1-r)
\epsilon_1+(1-s)\epsilon_2} \,, \ W = \sum_{A=1}^N e^{i\Sigma_A} \,, \ V^* = V|_{e\to e^{-1}} \,, \ W^* = W|_{e\to e^{-1}}
\end{equation}
and the character of the tangent space at such a fixed point is
\begin{equation}
\label{chistd}
\chi_{\lambda} = -V^*V(1-e^{\epsilon_1})(1-e^{\epsilon_2})+W^*V+V^*We^{\epsilon_1+\epsilon_2} \,,
\end{equation}
with positive contributions from $(A,B,P,Q)$ and negative ones from the
$GL(d,\mathbb{C})$ quotient and the relation $[A,B]+PQ=0$. After
cancellations, $\chi_{\lambda}=\sum_{l=1}^{2Nd}e^{x_l(\lambda)}$ and
the instanton partition function reads
\begin{equation}
Z = \sum_{\lambda} q^{\lvert\lambda\rvert} Z^{\text{4d}}_{
\text{vector}}(\lambda) \,,
\quad\text{with}\quad Z^{\text{4d}}_{
\text{vector}}(\lambda) = \prod_{l=1}^{2Nd}
\frac{1}{x_l(\lambda)} \,.
\end{equation}

For the case of a $\mathbb{Z}_M$ orbifold we rescale $\epsilon_2
\to\epsilon_2/M$ so that the orbifold acts by $\epsilon_2\to
\epsilon_2+2\pi i$. Indices ranging from $1$ to~$M$ are implicitly
taken modulo~$M$ when outside this range. We label colors by
$(j,a)$ with $1\leq j\leq M$ (their $\mathbb{Z}_M$ charge) and
$1\leq a\leq n_j$, and we shift the gauge equivariant parameters
$i\Sigma^j_a\to i\Sigma'^j_a=i\Sigma^j_a+j\epsilon_2/M$
so that they are invariant under the orbifold. The character of~$W$ is
then
\begin{equation}
W^{\mathbb{Z}_M} = \sum_{j=1}^M \sum_{a=1}^{n_j} e^{i\Sigma'^j_a-j\epsilon_2/M} \,.
\end{equation}

Fixed points are again labeled by $N$-tuples of Young diagrams~$
\lambda$ and the character of~$V$ is
\begin{equation}
V^{\mathbb{Z}_M} = \sum_{j=1}^M \sum_{a=1}^{n_j}
\sum_{(r,s)\in\lambda^j_a} e^{i\Sigma'^j_a-j\epsilon_2/M+(1-r)
\epsilon_1+(1-s)\epsilon_2/M} \,.
\end{equation}

The character $\chi^{\mathbb{Z}_M}_{\lambda}=\sum_l\exp x^{\mathbb{Z}_M}_l(\lambda)$ of the tangent space is obtained as the
orbifold-invariant terms (those with integer multiples of~$\epsilon_2$) in the standard character~\eqref{chistd} with $\epsilon_2
\to\epsilon_2/M$ and $\{i\Sigma\}\to\{i\Sigma'^j_a-j\epsilon_2/M\}$. The orbifolded instanton partition function involves
$M$~instanton counting parameters $q_j$ and reads
\begin{equation}
Z^{\mathbb{Z}_M}_{\text{instanton}} = \sum_{\lambda} Z^{\mathbb{Z}_M}_{\text{vector}}(\lambda) \biggl( \prod_{j=1}^M \prod_{a=1}^{n_j} \prod_{(r,s)\in\lambda^j_a} q_{j+s-1} \biggr)
\quad
\text{with}\quad Z^{\mathbb{Z}_M}_{\text{vector}}(\lambda) =
\prod_l\frac{1}{x^{\mathbb{Z}_M}_l(\lambda)} \,.
\end{equation}

The product of $q$ can be written as $\prod_{i=1}^M q_i^{d_i(\lambda)}$ in terms of the dimension
\begin{equation}\label{dioflambda}
d_i(\lambda) = \sum_{s=1}^\infty \sum_{a=1}^{n_{i-s+1}} (\widetilde{\lambda}^{(i-s+1)}_a)_s
\end{equation}
of the $\omega^i$~eigenspace $V_i\subset V$ of~$\mathbb{Z}_M$.

\subsection{Integral representation}

To simplify the comparison with Kanno--Tachikawa we shift $\phi
\to\phi-\frac{\epsilon_1+\epsilon_2}{2}$ compared to previous
sections, and the surface operator is along the $\epsilon_1$ plane
instead of $\epsilon_2$.

We can apply the orbifolding procedure to the integral representation
of the standard partition function of $d$ $U(N)$ instantons. We map
$\epsilon_2\to\epsilon_2/M$ and $\{i\Sigma\}\to\{i\Sigma'^j_a-j\epsilon_2/M\}$ and keep only the orbifold-invariant factors,
namely factors with integer multiples of~$\epsilon_2$. The
$\prod_j q_j^{d_j}$ term in the orbifolded instanton partition
function reads $Z^{\mathbb{Z}_M}(\vec{d})=\int d^d\phi\, Z^{
\mathbb{Z}_M}(\phi)$ with
\begin{multline}\label{orbifoldedintegral}
  Z^{\mathbb{Z}_M}(\phi) =
  \prod_{j=1}^M \Biggl[ \prodprime_{I,J=1}^{d_j} \frac{\phi^j_I-\phi^j_J}{\phi^j_I-\phi^j_J+\epsilon_1}
  \prod_{I=1}^{d_j} \prod_{J=1}^{d_{j-1}} \frac{\phi^j_I-\phi^{j-1}_J+\epsilon_2/M+\epsilon_1}{\phi^j_I-\phi^{j-1}_J+\epsilon_2/M}
  \\
  \times \prod_{a=1}^{n_j} \frac{1}{\prod_{I=1}^{d_j}(\phi^j_I-i\Sigma'^j_a+j\epsilon_2/M) \prod_{I=1}^{d_{j-1}}(i\Sigma'^j_a-\phi^{j-1}_I+\epsilon_1-(j-1)\epsilon_2/M)} \Biggr]
\end{multline}
and the Kanno--Tachikawa expression is a sum of residues at
\begin{equation}\label{phiKTpole}
\phi= \{i\Sigma'^i_a+(1-r)\epsilon_1+(1-s-i)\epsilon_2/M
\mid1\leq i\leq M,1\leq a\leq n_i,(r,s)\in\lambda^i_a\} \,,
\end{equation}
more precisely $\phi^j$ is the set of these values for $i+s-1\equiv
j\bmod{M}$. This contour prescription is equivalent to the
$\sum\phi$ JK prescription. The proof is essentially identical to the
case without orbifold: orbifolding only removes factors that have
non-zero coefficient of $\epsilon_2/M$ hence that are non-zero.

Let us split the $\sum\phi$ JK prescription in two steps, performing
first the $\phi^1,\ldots,\phi^{M-1}$ integrals and afterwards the
$\phi^M$ integrals:
\begin{equation}
\int d^{d_M}\!\phi^M \biggl( \int d^{d-d_M}\!\phi\, Z^{
\mathbb{Z}_M}(\phi) \biggr)
= \int d^{d_M}\!\phi^M I(\phi^M,i
\Sigma') = \sum_{\lvert\mu\rvert=d_M} \mathop{\mathrm{res}}_{
\phi^M=\phi^M(\mu)} I(\phi^M,i\Sigma') \,.
\end{equation}

The integrand $I(\phi^M,i\Sigma')$ obtained from the first step
consists of ratios of factors linear in $\phi^M$ and $i\Sigma'$, with
poles at values of $\phi^M$ deduced from~\eqref{phiKTpole}, namely
\begin{equation}
\phi^M(\mu)=\{i\Sigma'^i_a+(1-r)\epsilon_1-t\epsilon_2
\mid1\leq i\leq M,1\leq a\leq n_i,(r,t)\in\mu^i_a\}
\end{equation}
in terms of the $N$ Young diagrams
\begin{equation}
\mu^i_a=\{(r,t)\in\mathbb{Z}_{\geq1}^2\mid(r,tM+1-i)\in
\lambda^i_a\} \,.
\end{equation}

From the point of view of the $\phi^{1,\ldots,M-1}$ integrals, the
$\phi^M$~singularities arise when some of the $\phi^{1,\ldots,M-1}$
contours are pinched, namely when poles that lie on different sides of
the contour collide. The $\phi^M$~residue is then the residue at
these colliding poles as in the following toy model. Consider $\int
dx\,f(x)/[(a-x)(x-b)]=f(b)/(a-b)$ for $f$~smooth, with a contour
for~$x$ that selects the pole at~$b$; the result is singular at $a=b$
because the $x$~contour is pinched between the poles at $a$ (not
selected) and~$b$ (selected); the residue at $a=b$ is found as the
residue at $a=x=b$ in the original integral. The residue at a pole
$\phi^M=\phi^M(\mu)$ is similarly obtained by taking residues for
some of the $\phi^{1,\ldots,M-1}$ too. We denote these constrained~$
\phi^j$ (including~$\phi^M$) by~$\phi^{\prime\,j}$ and the
remaining variables by~$\phi''^j$. Explicitly,
\begin{equation}
\phi'(\mu) = \{i\Sigma^i_a+(1-r)\epsilon_1+(1-s)\epsilon_2/M
\mid1\leq i\leq M,1\leq a\leq n_i,(r,s)\in\lambda^{\prime\,
i}_a\}
\end{equation}
in terms of $i\Sigma^i_a=i\Sigma'^i_a-i\epsilon_2/M$ and of
the Young diagrams
\begin{equation}
\lambda^{\prime\,i}_a=\left\{ (r,s)\in\mathbb{Z}_{\geq1}^2
\middle|
\bigl(r,\lceil(s+i-1)/M\rceil\bigr)\in\mu^i_a\right\}
\subseteq
\lambda^i_a \,.
\end{equation}
As usual, $\phi'(\mu)$ splits into $\phi^{\prime\,j}(\mu)$
according to
$j\equiv i+s-1\bmod{M}$.

The (partial) residue $Z^{\mathbb{Z}_M}(\mu,\phi'')$ of
$Z^{\mathbb{Z}_M}(\phi)$ at $\phi'=\phi'(\mu)$ is given by the
orbifold-invariant factors (with integer multiples of~$\epsilon_2$)
in the corresponding residue $Z(\lambda',\phi'')$ of the standard
instanton partition function with $\epsilon_2\to\epsilon_2/M$ and
$\phi\to\phi-\frac{\epsilon_1+\epsilon_2}{2}$. We compute
\begin{multline}\label{partialresidue}
  Z(\lambda',\phi'') = Z(\lambda')
  \prod_J \prod_{i,a} \frac{\prod_{(r,s)\in\partial_-\lambda'^i_a}(i\Sigma^i_a-(r-1)\epsilon_1-(s-1)\epsilon_2/M-\phi''_J)} {\prod_{(r,s)\in\partial_+\lambda'^i_a}(i\Sigma^i_a-(r-2)\epsilon_1-(s-2)\epsilon_2/M-\phi''_J)}\\
  \times \prod_J \prod_{i,a} \frac{\prod_{(r,s)\in\partial_-\lambda'^i_a}(\phi''_J-i\Sigma^i_a+r\epsilon_1+s\epsilon_2/M)} {\prod_{(r,s)\in\partial_+\lambda'^i_a}(\phi''_J-i\Sigma^i_a+(r-1)\epsilon_1+(s-1)\epsilon_2/M)}
  \prodprime_{I,J} \frac{\phi''_{IJ}(\phi''_{IJ}+\epsilon_1+\epsilon_2/M)}{(\phi''_{IJ}+\epsilon_1)(\phi''_{IJ}+\epsilon_2/M)}
\end{multline}
where $\prod_{i,a}$ is $\prod_{i=1}^M\prod_{a=1}^{n_i}$ and
\begin{equation}\label{zlambdaprime}
Z(\lambda') = \mathop{\mathrm{res}}_{\phi'=\phi'(\lambda')}\Biggl
[ \frac{\prodprime_{I,J} \frac{\phi'_{IJ}(\phi'_{IJ}+\epsilon_1+
\epsilon_2/M)}{(\phi'_{IJ}+\epsilon_1)(\phi'_{IJ}+\epsilon_2/M)}}{
\prod_I \prod_{A=1}^N(\phi'_I-i\Sigma_A)(-\phi
'_I+i\Sigma_A+\epsilon_1+\epsilon_2/M)} \Biggr]
\end{equation}
is the usual contribution to the Nekrasov partition function. Given the
construction of~$\lambda'$, we have $(\lambda^{\prime\,i}_a)_r=\max(0,M( \mu^i_a)_r+1-i)$ hence
\begin{gather}
  \partial_-\lambda'^i_a = \left\{(r,tM+1-i)\middle|(r,t)\in\partial_-\mu^i_a\right\}\\
  \label{partialpluslambdaprime}
  \partial_+\lambda'^i_a = \left\{\begin{matrix}(r,1)\text{ if }t=1,\\(r,tM-M+2-i)\text{ if }t>1\end{matrix}\middle|(r,t)\in\partial_+\mu^i_a\right\} \,.
\end{gather}

We can now perform the $\phi''$ integrals using the $\sum\phi''$ or
$-\sum\phi''$ JK prescriptions. The first reproduces the set of poles
we already saw in~\eqref{phiKTpole}, since this is simply a long-winded
way of writing the $\sum\phi$ JK prescription. The effect of
orbifolding is to select poles that have the desired numbers~$d_j$ of
charge~$j$ components $\phi^j$ in~$\phi$, then for each residue to
keep only factors that have an integer multiple of~$\epsilon_2$.

\subsection{Flipping the contour}

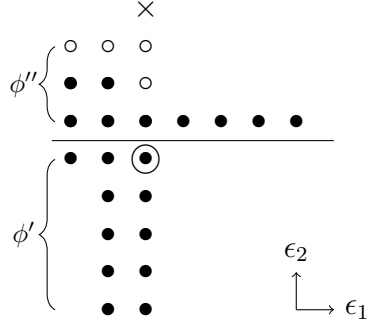
\begin{figure}
\centering
  \def\M{4}
  \def\I{4}
  \def\MU{1/2,2/2,3/1}
  \def\K{2/2,3/7}
  \begin{tikzpicture}[scale=.5]
    \def\rmax{3}
    \pgfmathsetmacro\SMAX{\M*2}
    \begin{scope}[shift={(0,0)}]
      \foreach \r/\smax in \MU {
        \pgfmathsetmacro \smaxtimesM {\smax*\M}
        \foreach \s in {\I,...,\smaxtimesM} {
          \node at (-\r,-\s) {$\bullet$};}}
      \pgfmathsetmacro\x{-\rmax-.5}
      \pgfmathsetmacro\y{-\I+.5}
      \draw (\x,\y) -- (4,\y);
      \pgfmathsetmacro\Iminusone{\I-1}
      \foreach \r in {1,...,\rmax} {
        \foreach \s in {1,...,\Iminusone} {
          \node at (-\r,-\s) {$\circ$};}}
      \foreach \s/\ks in \K {
        \foreach \r in {1,...,\ks} {
          \pgfmathsetmacro\x{\r-\rmax-1}
          \node at (\x,-\s) {$\bullet$};}}
      \node at (-1,-\I) {$\bigcirc$};
      \node at (-1,0) {$\times$};
      \draw [decorate,decoration={brace,amplitude=6pt},xshift=-1em] (-\rmax,-\Iminusone) -- (-\rmax,-1) node [midway,xshift=-1em]{\footnotesize $\phi''$};
      \draw [decorate,decoration={brace,amplitude=6pt},xshift=-1em] (-\rmax,-\SMAX) -- (-\rmax,-\I) node [midway,xshift=-1em]{\footnotesize $\phi'$};
    \end{scope}
    \begin{scope}[shift={(3,-8)}]
      \draw [->] (0,0) -- (1,0) node [right] {$\epsilon_1$};
      \draw [->] (0,0) -- (0,1) node [above] {$\epsilon_2$};
    \end{scope}
  \end{tikzpicture}
  \newcommand{\drawmu}{\smash{$\!\vcenter{\hbox{\tikz[scale=.3]\foreach \r/\smax in \MU \foreach \s in {1,...,\smax} \node at (\r,\s) {$\bullet$};}}\!$}}
  \newcommand{\drawk}{\smash{$\!\vcenter{\hbox{\tikz[scale=.3]\foreach \s/\rmax in \K \foreach \r in {1,...,\rmax} \node at (\r,-\s) {$\bullet$};}}\!$}}
\caption{Components of~$\phi'$ corresponding to a given choice of $1\leq i\leq
M$ and $1\leq a\leq n_i$ (here $M=4$ and $i=4$) are described by a Young
diagram $\lambda^{\prime\,i}_a$ deduced from $\mu^i_a$ (here
\protect\drawmu) by subtracting~$(i-1)$ from $M$~times the lengths of columns.
Components of~$\phi''$ are described by $\lambda''^i_a$ (here
\protect\drawk) with at most $i-1$ rows. The hollow bullets $\circ$ indicate
when some of the $k_{ias}$ are negative (here $k_{ia1}=-3$, $k_{ia2}=-1$,
$k_{ia3}=4$). The circled bullet  $\ \mathclap{{\bigcirc}}\mathclap{{\bullet}}\ $ is at~$i \Sigma^i_a$. The cross $\times$ is at~$i\Sigma'^i_a$.}
\label{fig:examplepole}
\end{figure}

The $-\sum\phi''$ JK prescription\footnote{This two-step procedure
integrating $\phi'$ then~$\phi''$ is in fact equivalent to the
$-\sum\phi^1-\cdots-\sum\phi^{M-1}+\xi\sum\phi^M$ JK
prescription with $\xi\gg1$.} selects poles at $\phi''_J=i\Sigma^i_a+(2-r)\epsilon_1+(2-s)\epsilon_2/M$ for $(r,s)\in
\partial_+\lambda^{\prime\,i}_a$ and at shifts thereof due to the usual
$\phi''_J=\phi''_I+\epsilon_1$ and $\phi''_J=\phi''_I+
\epsilon_2/M$ for $I<J$. Each box in~\eqref{partialpluslambdaprime}
with $t>1$ leads to a pole $\phi''_j=i\Sigma'^i_a+(2-r)
\epsilon_1+(1-t)\epsilon_2$ with $\mathbb{Z}_M$~charge~$M$. By
construction, $\phi''^M=\emptyset$ hence these poles do not
contribute to the orbifolded partition function. Poles are thus at
\begin{equation}\label{phiprimeprime}
\phi''=\left\{ i\Sigma^i_a+(r-(\widetilde{\mu}^i_a)_1)
\epsilon_1+s\epsilon_2/M\middle|(r,s)\in\lambda''^i_a\right\}
\,,
\end{equation}
based on the pole $i\Sigma^i_a+(1-(\widetilde{\mu}^i_a)_1)
\epsilon_1+\epsilon_2/M$ corresponding to $t=1$ in $\partial_+
\mu^i_a$. The $N$~Young diagrams $\lambda''^i_a$ are
due as
usual to the $\phi''_J=\phi''_I+\epsilon_1$ and $\phi''_J=
\phi''_I+\epsilon_2/M$ poles. The constraint $\phi''^M=\emptyset
$ implies that $\lambda''^i_a$ has at most $(i-1)$ rows.
The data
of $\lambda''^i_a$ can be recast as integers
\begin{equation}
k_{ias} = \bigl(\widetilde{\lambda''}^i_a\bigr)_{i-s} - (
\widetilde{\mu}^i_a)_1 \geq-(\widetilde{\mu}^i_a)_1
\quad\text{for $1\leq s<i\leq M$ and $1\leq a\leq n_i$}\,,
\end{equation}
with $k_{ias}\leq k_{ia(s+1)}$, and we will soon reinterpret these as
2d vorticities. See \autoref{fig:examplepole} for a depiction of
components of~$\phi$.

Let us compute $Z^{\mathbb{Z}_M}(\mu,k)$, namely orbifold-invariant
factors in the residue $Z(\lambda',\lambda'')$ of $Z(\lambda',
\phi'')$ at~\eqref{phiprimeprime}. This residue is simply the residue
at $\phi=\phi'(\lambda')\cup\phi''(\lambda'')$ of the standard
instanton partition function integral (with $\epsilon_2\to\epsilon_2/M$\ldots{}). We first perform product over~$I$,
\begin{align}
  & Z(\lambda',\lambda'') \nonumber\\
  & = \res \biggl[ \prodprime_{I,J=1}^d \frac{\phi_{IJ}(\phi_{IJ}+\epsilon_1+\epsilon_2/M)}{(\phi_{IJ}+\epsilon_1)(\phi_{IJ}+\epsilon_2/M)} \prod_{J=1}^d\prod_{A=1}^N\frac{1}{(\phi_J-i\Sigma_A)(-\phi_J+\epsilon_1+\epsilon_2/M+i\Sigma_A)} \biggr]
  \nonumber\\\label{zlplpp}
  & \begin{multlined}
    = \prod_{i=1}^M\prod_{a=1}^{n_i} \prod_{J=1}^d \biggl[ \prod_{u=1}^{i-1} \frac{i\Sigma'^i_a+(1+k_{iau})\epsilon_1-(u-1)\epsilon_2/M-\phi_J}{i\Sigma'^i_a+(1+k_{iau})\epsilon_1-u\epsilon_2/M-\phi_J} \\
    \times \frac{\prod_{(r,s)\in\partial_-\mu^i_a} (i\Sigma'^i_a-(r-1)\epsilon_1-s\epsilon_2-\phi_J)} {(\phi_J-i\Sigma'^i_a+(i/M)\epsilon_2)\prod_{(r,s)\in\partial_+\mu^i_a}(i\Sigma'^i_a-(r-2)\epsilon_1-(s-1)\epsilon_2-\phi_J)} \biggr]
  \end{multlined}
\end{align}
then we use
\begin{multline}
  \prod_{J=1}^d (\phi_J-\alpha) = \prod_{i=1}^M \prod_{a=1}^{n_i} \biggl[ \prod_{(r,s)\in\mu^i_a} \prod_{j=1}^M \biggl(i\Sigma'^i_a-(r-1)\epsilon_1-\Bigl(s-1+\frac{j}{M}\Bigr)\epsilon_2-\alpha\biggr) \\
  \prod_{u=1}^{i-1} \biggl( \Bigl(\frac{i\Sigma'^i_a-u\epsilon_2/M-\alpha}{\epsilon_1}+1\Bigr)_{k_{iau}} \epsilon_1^{k_{iau}} \biggr) \biggr] \,,
\end{multline}
where $(\beta)_k=\Gamma(\beta+k)/\Gamma(\beta)$ is the Pochhammer symbol.

We only keep orbifold-invariant factors and eventually we get
\begin{equation}\label{zorbi}
Z^{\mathbb{Z}_M}_{\text{instanton}}(i\Sigma,q) = \frac{1}{Z_0}
\sum_{\mu,k} \Bigl[ Z^{\text{4d}}(i\Sigma',\Phi;q) Z^{\text{4d--2d}}(
\Phi,i\sigma_1) Z^{\text{2d}}(i\Sigma',i\sigma;z) \Bigr]
\end{equation}
where $q=\prod_{j=1}^M q_j$ and $z_j=q_j/\epsilon_1^{n_j+n_{j+1}}$ (using $d_M=\lvert\mu\rvert$ and $d_s=d_M+\sum_{i=s+1}^M \sum_{a=1}^{n_i} k_{ias}$), $Z^{\text{4d}}$ is as in \eqref{instantonvortexpieces} but with $M=M'=0$ in that formula's
notations,
\begin{align}
Z^{\text{4d--2d}}(\Phi,i\sigma_1)
& = \prod_{J=1}^{d_M} \prod_{p=1}^{r_1} \frac{i\sigma_{1,p}-\Phi_J+
\frac{\epsilon_1+\epsilon_2}{2}}{i\sigma_{1,p}-\Phi_J+\frac{
\epsilon_1-\epsilon_2}{2}} \,,
\\
Z^{\text{2d}}(i\Sigma',i\sigma;z)
& =
\prod_{j=1}^{M-1} \frac{((-1)^{n_j+n_{j+1}}z_j)^{\Tr(i\sigma_j/\epsilon_1)}
\prod_{p,q=1}^{r_j} \Gamma(1+(i\sigma_{j,p}-i\sigma_{j,q})/\epsilon_1)}{\prod_{p=1}^{r_j} \prod_{q=1}^{r_{j-1}} \Gamma(1+(i\sigma_{j,p}-i\sigma_{j-1,q})/\epsilon_1)} \,,
\end{align}
in terms of $r_j = \sum_{i=j+1}^M n_i$,
\begin{equation}\compressthisequation
  \Phi=\phi^M+\tfrac{\epsilon_2-\epsilon_1}{2}=\{i\Sigma'^i_a+(\tfrac{1}{2}-t)\epsilon_1+(\tfrac{1}{2}-u)\epsilon_2|1\leq i\leq M,1\leq a\leq n_i,(t,u)\in\mu^i_a\}\,,
\end{equation}
\begin{equation}
  \{i\sigma_{j,p}\mid 1\leq p\leq r_j\}=\{i\Sigma'^i_a+k_{iaj}\epsilon_1\mid j<i\leq M,1\leq a\leq n_i\} \text{ for $0\leq j<M$}\,,
\end{equation}
and in particular $\{i\sigma_{0,p}\}=\{i\Sigma'^i_a\}$. The
normalization $Z_0=Z^{\text{2d}}|_{k=0}$ simply sets the $\mu=
\emptyset$, $k=0$ coefficient to~$1$.

\subsection{4d--2d description and dualities}
\begin{figure}
\centering
  \begin{tikzpicture}
    \tikzset{->-/.style 2 args={decoration={markings, mark=at position #1 with {\arrow[rotate=#2]{stealth}}},postaction={decorate}}}
    \node[draw,circle] (N) at (0,0) {$N$};
    \node[draw,circle] (r1) at (1.5,0) {$r_1$};
    \node[draw,circle] (r2) at (3,0) {$r_2$};
    \node (dots) at (4.5,0) {$\cdots$};
    \node[draw,circle,inner sep=1pt,minimum size=6ex] (rM1) at (6,0) {$\mathclap{r_{M-1}}$};
    \draw[->-={.65}{0}] (N) -- (r1);
    \draw[->-={.65}{0}] (r1) -- (r2);
    \draw[->-={.55}{0}] (r2) -- (dots);
    \draw[->-={.55}{0}] (dots) -- (rM1);
    \node at (-.9,0) {\small 4d};
    \draw[dashed,rounded corners] (.6,-.6) rectangle (-1.2,.6);
    \draw[dashed,rounded corners] (.6,-.6) rectangle (7.2,.6);
    \node at (6.9,0) {\small 2d};
    \begin{scope}[shift={(0,-1.5)}]
      \node[draw,circle] (N) at (0,0) {$N$};
      \node[draw,circle,inner sep=1pt,minimum size=7ex] (r1) at (1.5,0) {\footnotesize $\mathclap{N{-}n_M}$};
      \node (dots) at (3,0) {$\cdots$};
      \node[draw,circle,inner sep=1pt,minimum size=6ex] (rM2) at (4.5,0) {\footnotesize $\mathclap{n_1{+}n_2}$};
      \node[draw,circle] (rM1) at (6,0) {\footnotesize $n_1$};
      \draw[->-={.45}{0}] (r1) -- (N);
      \draw[->-={.55}{0}] (dots) -- (r1);
      \draw[->-={.55}{0}] (rM2) -- (dots);
      \draw[->-={.55}{0}] (rM1) -- (rM2);
      \node at (-.9,0) {\small 4d};
      \draw[dashed,rounded corners] (.6,-.7) rectangle (-1.2,.7);
      \draw[dashed,rounded corners] (.6,-.7) rectangle (7.2,.7);
      \node at (6.9,0) {\small 2d};
    \end{scope}
  \end{tikzpicture}
\caption{Two dual quiver descriptions of the Gukov--Witten surface operator that
breaks $SU(N)$ to $S[U(n_1)\times\cdots\times U(n_M)]$ by imposing a
monodromy~\eqref{GWmonodromy} $A\sim i\alpha d \theta$ where $\alpha$
has $n_j$~eigenvalues $\alpha_{(j)}$, such that
$0\leq\alpha_{(M)}<\cdots<\alpha_{(1)}<2\pi$. The quiver description at the top
(which we use the most) is in terms of a 2d $\Nsusy=(2,2)$ theory with
gauge group ranks $r_s=n_{s+1}+ \cdots+n_M$ and bifundamental chiral
multiplets. The bottom quiver description has gauge group ranks
$n_1+\cdots+n_s$ for $M-1\geq s\geq1$. In both case, the $SU(N)$ flavor
symmetry of the 2d theory is gauged by the 4d gauge group and there is no
superpotential.}
\label{fig:orbifold-quiver}
\end{figure}
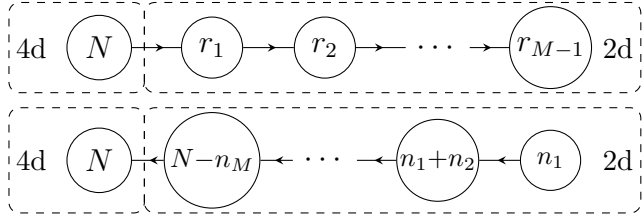

The Gukov--Witten surface operator admits a description as the 4d--2d
quiver gauge theory depicted in \autoref{fig:orbifold-quiver}. The 2d
$\Nsusy=(2,2)$ theory has gauge group $\prod_{i=1}^{M-1} U(r_i)$ with $r_i=n_{i+1}+\cdots+n_M$ and has chiral multiplets in
the $\bigoplus_{i=1}^{M-1} (\overline{r_{i-1}}\otimes r_i)$
representation of this gauge group and of an $SU(r_0)=SU(N)$ flavor
symmetry group, itself gauged using the 4d $\Nsusy=2$ vector
multiplet. Our prescription for the instanton--vortex partition function
of this quiver precisely reproduces~\eqref{zorbi}.

Indeed,
\begin{itemize}
\item $Z^{\text{4d}}(i\Sigma',\Phi)$ is the usual contribution of an $N$-tuple of Young diagrams to the instanton partition function without surface operator, with gauge equivariant parameters $\{i\Sigma'^i_a\}$ and instanton counting parameter $q=\prod_{j=1}^M q_j$;
\item $Z^{\text{4d--2d}}(\Phi,i\sigma)$ is due to the matter charged under both 4d and 2d gauge groups, with $\epsilon_1\leftrightarrow \epsilon_2$ compared to~\eqref{instantonvortexpieces} because the surface operator is in the $\epsilon_1$~plane, and opposite signs of $i\sigma_1$ and $\Phi_J$ because the bifundamental is in the conjugate representation;
\item $Z^{\text{2d}}(i\Sigma',i\sigma;z)$ is the usual contribution to the vortex partition function of the 2d theory, in which mass parameters of the $\overline{N}\otimes r_1$ chiral multiplet are the 4d gauge equivariant parameters $i\sigma_0=\{i\Sigma'^i_a\}$.
\end{itemize}
The renormalized FI parameter~$\zeta_j$ and theta angle~$\vartheta_j$ of each 2d gauge group are given by $z_j=e^{-2\pi\zeta_j+i
\vartheta_j}=q_j/\epsilon_1^{n_j+n_{j+1}}$. The power of
$\epsilon_1$ is explained for real~$\epsilon_1$ as a logarithmic
renormalization of the FI parameter, with coefficient equal to the
difference $r_{j-1}-r_{j+1}=n_j+n_{j+1}$ between the number of
fundamental and antifundamental chiral multiplets. The exponentiated FI
parameter~$z_j^{\text{UV}}$ at appearing in the UV Lagrangian
description is $z_j^{\text{UV}}=q_j/M_{\text{UV}}^{n_j+n_{j+1}}
\gg z_j$ hence we can assume $\lvert z_j^{
\text{UV}}\rvert<1$ namely $\zeta_j^{\text{UV}}>0$ later on for this
theory.

The 2d quiver gauge theory admits a web of dual descriptions generated
by Seiberg-like dualities. Some of these are obtained by repeatedly
using the IR equivalence between a $U(k)$ gauge theory with $U(n)$
flavor symmetry and $k\otimes\overline{n}$ chiral multiplets and a
$U(n-k)$ gauge theory with $\overline{n-k}\otimes n$ chiral multiplets.
Let us apply $(M-1)(M-2)/2$ of these elementary Seiberg-like dualities
namely applying them in turn to $M-1$ nodes starting from the
right-most $U(r_{M-1})$ then to $M-2$ nodes starting again from the
right-most, and so on until applying one duality to the right-most
gauge group. The result is the second quiver depicted in
\autoref{fig:orbifold-quiver}, namely a $\prod_{j=1}^{M-1} U(r'_j)$
gauge theory where $r'_j=n_1+\cdots+n_j$ with chiral multiplets
in $\bigoplus_{j=1}^{M-1}(r'_{j+1}\otimes\overline{r'_j})$. The
exponentiated FI parameter for the $U(r'_j)$ gauge group works out to
be $z'_j=1/z_j$.

This dual description is expected: under charge conjugation the order
of parameters $\alpha_{(j)}$ of the monodromy~\eqref{GWmonodromy}
defining the Gukov--Witten surface operator is reversed. The map of FI
parameters can be understood classically by realizing FI parameters as
differences between consecutive~$\alpha_{(j)}$. For this, we study
classical vacua of the 2d theory. These are solutions of the D-term
equations
\begin{equation}
Q_jQ_j^{\dagger}=\zeta_j^{\text{UV}} \operatorname{id}_{r_j}+Q_{j+1}^{\dagger}Q_{j+1} \quad\text{for } 1\leq j\leq M-1
\end{equation}
in terms of the chiral multiplets $Q_j\in\overline{r_{j-1}}\otimes
r_j$ and $Q_M=0$. Since $\zeta_j^{\text{UV}}>0$ the right-hand
side is a positive-definite matrix hence has full rank,~$r_j$. The matrix~$Q_j$ must thus have full rank too, hence it defines an embedding
$\mathbb{C}^{r_j}\hookrightarrow\mathbb{C}^{r_{j-1}}$. A solution of
the D-term equations is thus equivalent to a choice of flag
$\mathbb{C}^{r_{M-1}}\hookrightarrow\cdots\hookrightarrow
\mathbb{C}^{r_1}\hookrightarrow\mathbb{C}^N$. Integrating out 2d
matter gives rise to a source term $-Q_1^{\dagger}Q_1$ localized
on the surface for the 4d gauge field strength, and this source has
eigenvalues
\begin{equation}
\alpha^{\text{classical}}_{(j)}=-\zeta_1^{\text{UV}}-\cdots
-\zeta_{j-1}^{\text{UV}} \text{ with multiplicity }r_{j-1}-r_j=n_j\text{
for }1\leq j\leq M
\end{equation}
that reproduce the ordering $\alpha_{(M)}<\cdots<\alpha_{(1)}$. This
classical analysis is incomplete in an obvious way: the $\alpha_{(j)}$
are $2\pi$~periodic. However, one can hope that quantum effect do not
spoil the order.

One can cyclically permute the $\alpha_{(j)}$ by a gauge transformation,
together with the $n_j$ and the~$q_j$. How is this realized in the
4d--2d description? It is not simply a 2d duality because the 2d theory
on its own does not depend on~$q_M$. Instead, this duality only exists
thanks to the coupling to 4d, like the node-hopping duality
\cite{Gadde:2013ftv}. Note that the instanton partition
function~\eqref{orbifoldedintegral} is invariant under such permutations
keeping $i\Sigma^j_a=i\Sigma'^j_a-j\epsilon_2/M$
fixed, while
the quiver description involves $i\Sigma'^j_a$~instead.
Besides an
unimportant overall shift by $\epsilon_2/M$, the basic cyclic
permutation $1\to2\to\ldots\to M\to1$ also shifts $i\Sigma^{\prime
\,M}_a\to i\Sigma^{\prime\,M}_a-\epsilon_2$. Such a shift was also
found in
\cite{Gomis:2014eya} when studying node-hopping duality. It would
be interesting to understand its gauge theory origin better.

To prove the equality of the orbifolded-instanton and the
instanton--vortex partition functions, we compared two JK prescriptions.
Other JK prescriptions taking $K$ of the $\phi^j$ with coefficient
$+1$ and the $M-K$ other with coefficient $-\delta$ appear to give
instanton--vortex partition functions of 2d theories with $M-K-1$ gauge
groups, coupled to the 4d theory in the presence of a $\mathbb{Z}_K$
orbifold. This is consistent with the classical analysis we did above:
the orbifold is equivalent to a monodromy and integrating out 2d matter
breaks the 4d gauge group further by adding a source term in the
monodromy. It would be interesting to investigate dualities relating
these equivalent constructions of the GW surface operator.

It would also be interesting to compare our results with the modular anomaly equation discussed in \cite{ashok}.

\section{Conclusion}

In this paper we have clarified several aspects of the 2d/4d
correspondence and obtained the compact formula~\eqref{mainresult}
relating the 4d--2d partition function to the 2d nonperturbative counting
problem. In the 2d worldsheet theory we take into account the vortex and
negative vortex configurations. We have explicitly demonstrated that our
approach reproduces known 4d--2d partition functions for defects arising
from Higgsing and orbifolding.

It would be interesting to find how these findings are related to
various aspects of integrability. The most interesting point concerns
the precise identification of the geometric transition in the 4d
$\Omega$-background as the combination of several dualities known in
the integrability framework. To this aim elliptic versions of these
dualities are necessary which are not elaborated enough. The example of
the periodic Toda chain considered in \cite{surf} in the flat
undeformed space provides the simplest pattern of a classical--classical
duality between the Toda model and a particular limit of the spin chain
when the classical spectral curves are related. When the
$\Omega$-deformation is switched on, classical-quantum and
quantum--quantum dualities are expected in the elliptic models, and they
involve instanton contributions. We can conjecture that the geometric
transition amounts to different versions of dualities between the
systems of Calogero-RS type and the different spin chains. Certainly
this issue needs additional detailed study and we plan to discuss it
elsewhere.

Another interesting question to explore is that of intersecting
Gukov--Witten surface operators that break the gauge group to different
subgroups along the $\epsilon_1$ and $\epsilon_2$ planes. It is
natural to expect these to be equivalent to an orbifold $(\mathbb{C}/
\mathbb{Z}_L)\times(\mathbb{C}/\mathbb{Z}_M)$ where $\mathbb{Z}_L\times\mathbb{Z}_M$ acts as a gauge transformations too.
Following the same steps as for a single orbifold plane, one can write
the instanton partition function as an integral and change the JK
residue prescription. This leads to higher-order poles, hence we leave
that line of research to future work. The goal would be to find a
description of the intersecting operators as a 4d theory coupled to a
pair of 2d theories, each coupled to a matrix model on their
intersection.

\section*{Acknowledgement}
We are grateful to K.~Bulycheva, I.~Danilenko, N.~Nekrasov and
E.~Zenkevich for useful discussions. The research of A.G. was carried
out at the IITP RAS with the support of Russian Science Foundation grant
for IITP 14-50-00150. The work of N.S. was supported by RFBR grant 15-02-02092. 
A.M. is grateful to RFBR grant 15-02-02092 for travel
support. A.M. also thanks A. Alfieri for providing excellent food.

\appendix

\section{Nekrasov partition function}\label{a:nek}

In this section we briefly review the computation of Nekrasov partition function for $\Nsusy=2$ gauge theories.  We focus on A-type quiver gauge theories with a number of gauge factors $U(N_i)$, $1\leq i \leq V$, bifundamental hypermultiplets charged under $U(N_i) \times U(N_{i+1})$ for $0\leq i\leq V$, where $U(N_0)$ and $U(N_{V+1})$ are flavor symmetries.  The IIA construction is depicted in \autoref{fig:a_quiver}.

\renewcommand{\thefigure}{A.\arabic{figure}}

\begin{figure}
\centering
\begin{tikzpicture}[brane/.style={line width=1pt}, d4/.style={line width=2.4pt}]
  \draw[d4] (-1.5,1.3) -- (1,1.3);
  \draw[d4] (-1.5,2.3) -- (1,2.3);
  \draw[brane] (1,0) -- (1,3.7) node [below left] {NS5};
  \draw[d4] (1,1) -- (3,1);
  \draw[d4] (1,1.8) -- (3,1.8);
  \draw[d4] (1,3) -- (3,3) node [midway, above] {D4};
  \draw[densely dashed,{stealth}-{stealth},thick] (1.03,.4) -- (2.97,.4) node [midway, below] {$\Lambda_1$};
  \draw[brane] (3,0) -- (3,3.7);
  \draw[d4] (3,.6) -- (6,.6);
  \draw[d4] (3,1.4) -- (6,1.4);
  \draw[d4] (3,2.0) -- (6,2.0);
  \draw[d4] (3,2.4) -- (6,2.4);
  \draw[d4] (3,3.5) -- (6,3.5);
  \draw[densely dashed,{stealth}-{stealth},thick] (3.03,.3) -- (5.97,.3) node [midway, below] {$\Lambda_2$};
  \draw[brane] (6,0) -- (6,3.7);
  \draw[d4] (6,0.9) -- (8.5,0.9);
  \draw[d4] (6,2.8) -- (8.5,2.8);
  \draw[densely dashed,{stealth}-{stealth},thick] (6.03,.4) -- (8.47,.4) node [midway, below] {$\Lambda_3$};
  \draw[brane] (8.5,0) -- (8.5,3.7);
  \draw[d4] (8.5,1.7) -- (11,1.7);
\end{tikzpicture}
\caption{\label{fig:a_quiver}Brane construction of a $U(3) \times U(5) \times U(2)$ gauge theory.  The leftmost $U(3)$ factor has two fundamental hypermultiplets and the rightmost $U(2)$ factor has one fundamental hypermultiplet.}
\end{figure}
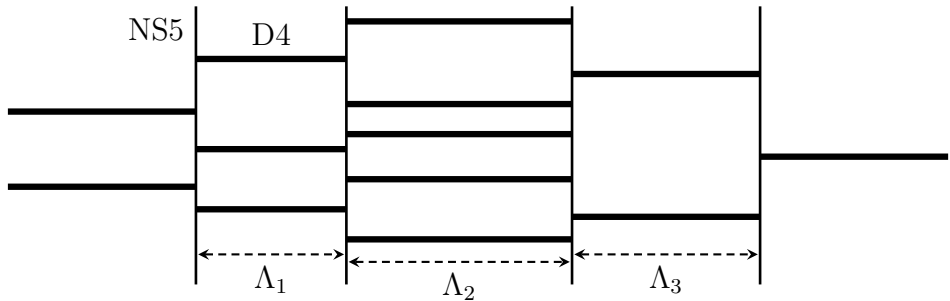

Each gauge factor has is own coupling constant and correspondingly the dynamical scale $\Lambda_i$.  We will be considering these theories in the Coulomb phase, parametrized by the Higgs scalar VEV $a^i=\{ a^i_1,\dots,a^i_{N_i} \}$ for $1\leq i\leq V$.  In \autoref{sec:coupling4dto2d} and \autoref{sec:Higgs} we denote $a^i=i\Sigma^i+\frac{\epsilon_1+\epsilon_2}{2}$ while in \autoref{sec:orbifold} it is more convenient to denote $a^i=i\Sigma^i$.  When there are only one or two gauge groups we write $\Sigma$, $\Sigma'$ instead of $\Sigma^1$, $\Sigma^2$.

The Nekrasov partition function is the equivariant volume of the instanton moduli space.  In the IIA construction, instantons are realized as D0 branes and the instanton moduli space is the vacuum moduli space of their world-volume theory.  Its equivariant volume is the matrix model integral
\begin{multline}
  Z^{\text{4d}}_{\text{instanton}} = \int \biggl[ \prod_{i=1}^V \frac{d^{K_i}\phi_i}{K_i!} \biggr] \frac{\prod_{i=1}^{V-1} \prod_{I=1}^{K_i} \prod_{J=1}^{K_{i+1}}S(\phi^{i+1}_J-\phi^i_I)}{\prod_{i=1}^V \bigl[ (\frac{\epsilon_1\epsilon_2}{\epsilon_1+\epsilon_2})^{K_i}\prod_{1\leq I\neq J\leq K_i} S(\phi^i_I-\phi^i_J)\bigr]}\\
  \times \prod_{i=1}^{V} \prod_{I=1}^{K_i} \frac{\prod_{A=1}^{N_{i-1}} (\phi^i_I-a^{i-1}_A+\epsilon_1+\epsilon_2) \prod_{A=1}^{N_{i+1}} (a^{i+1}_A-\phi^i_I)} {\prod_{A=1}^{N_i} [ (\phi^i_I-a^i_A+\epsilon_1+\epsilon_2)(a^i_A-\phi^i_I) ]}
\end{multline}
where we ignore powers of $2\pi i$ and where
\begin{equation}
S(\phi)=-\frac{(\phi+\epsilon_1)(\phi+\epsilon_2)}{\phi(-
\phi-\epsilon_1-\epsilon_2)} \,.
\end{equation}

In the IIA construction $S(\phi)$ factors captures the interaction of D0 branes on opposite sides of an NS5 brane, while D0 branes in the same interval between NS5 branes contribute the $1/S(\phi)$ factors.  The second line in the formula involves interactions between D0 and D4 branes in the same interval (denominator) or neighboring intervals (numerators).  The odd placement of signs in $S(\phi)$ is important in the next section.

The contour is chosen so as to enclose poles at
\begin{equation}\label{expectedsetofpoles}
\{\phi^i_I\mid1\leq I\leq K_i\} = \{a^i_A+(r-1)\epsilon_1+(s-1)\epsilon_2\mid(r,s)\in\lambda^i_A\}
\end{equation}
for each choice of Young diagrams $\lambda^i_A$, $1\leq i\leq V$, $1\leq A\leq N_i$ with $\sum_{A=1}^{N_i} \lvert\lambda^i_A \rvert=K_i$.  The freedom of ordering the $\phi^i_I$ cancels the $K_i!$ factors.

\subsection{Jeffrey--Kirwan residue prescription}

We prove here that the poles above are those selected by the Jeffrey--Kirwan (JK) residue prescription with parameter $\eta=- \sum_{i,I}\phi^i_I$.  We assume $\epsilon_1\neq\epsilon_2$ and that the $a^i_A$ are generic: other cases are obtained by continuity.  The proof proceeds as follows: we perturb denominators in the integral slightly to avoid higher-order poles, we write the result as a sum over colored trees, equivalent to closing contours one by one, and we then remove the perturbation and determine which poles remain in that limit.

By definition the JK prescription selects poles~$p$ for which there exists a set $\mathcal{D}$ of $\sum_i K_i$ denominator factors vanishing at~$p$ such that $\eta\in\operatorname{Cone}(\mathcal{D}_{\phi})$, namely $\eta$ is in the cone (is a linear combination with positive coefficients) of the $\phi$-dependent parts of the chosen denominators.  In our case, the $\phi$-dependent part of each denominator is $\pm\phi^i_I$ or $\phi^j_J-\phi^i_I$.  Consider one such choice $\mathcal{D}$ of denominators and define a graph $\Gamma_{\mathcal{D}}$ with vertices $\{(i,I)\mid1\leq i\leq V,1 \leq I\leq K_i\}\cup\{0\}$, an edge $(j,J)\to(i,I)$ for each $\phi^i_I-\phi^j_J+\dots\in\mathcal{D}$, an edge $(j,J) \to0$ for each $-\phi^j_J+\cdots\in\mathcal{D}$ and an edge $0\to(i,I)$ for $\phi^i_I+\cdots\in\mathcal{D}$.  Given the choice of $\eta$, every $\phi^i_I$ must appear at least once with a negative sign in $\mathcal{D}$, hence every vertex of $ \Gamma_{\mathcal{D}}$ except $0$ must have an outgoing edge.  Since there are one fewer edges than vertices the graph is a directed tree with root~$0$ (and in particular no $0\to(i,I)$ edge).  This tree only encodes the $\phi$-dependent part of denominators in $\mathcal{D}$ and it should be supplemented by coloring its edges
\begin{equation}
\begin{cases}
(i,J)\to(i,I)
& \text{by } \epsilon_j\ (\mbox{for }j=1,2)\mbox{ if
} \phi^i_I-\phi^i_J+\epsilon_j\in\mathcal{D}
\\
(i,I)\to0
& \text{by } A \text{ if } a^i_A-\phi^i_I\in
\mathcal{D} \,.
\end{cases}
\end{equation}
It is not necessary to color edges $(j,J)\to(i,I)$ for $i\neq j$ (more precisely $i=j\pm 1$) because there is only a single denominator with this $\phi$-dependent part.

For each choice of $\mathcal{D}$, the hyperplanes $h(\phi)=0$ for $h(\phi)\in\mathcal{D}$ intersect at a single point~$p_{\mathcal{D}}$ (equations can be solved starting from the root of $\Gamma_{ \mathcal{D}}$).  We would now like to compute the JK residue at~$p_{ \mathcal{D}}$, but typically many denominators apart from those in $\mathcal{D}$ vanish at~$p_{\mathcal{D}}$.  Correspondingly, many different~$ \mathcal{D}$ give the same $p_{\mathcal{D}}=p$.  The definition of JK residue instructs us to resolve this singularity by decomposing the integrand near~$p$ into elementary fractions with $\sum_i K_i$ denominator factors each, then sum residues of fractions for which $\eta$ belongs to the cone.  An often simpler approach is to perturb the problem slightly by adding to each denominator a different generic small parameter~$\delta_{\alpha}$, then to compute the JK residue without encountering any issue of higher-order pole, then to send all $\delta_{\alpha}\to0$.  The perturbation does not affect the set of $\mathcal{D}$ we have already worked out.  While the contribution of each $\mathcal{D}$ can blow up as $\delta_{\alpha}\to0$ the sum over $\mathcal{D}$ keeps a finite value, unless the contour is pinched.  Such a pinching may arise when a pole $p_{\mathcal{D}}$ selected by the JK prescription collides as $\delta_{\alpha}\to0$ with a pole that is not selected.  Thankfully, in all cases of relevance to us the contribution of each $\mathcal{D}$ remains finite.

We choose the following non-generic shift of denominators:
\begin{multline}\label{tweakedz4dintegral}
  \tilde{Z}^{\text{4d}} = \int \biggl[ \prod_{i=1}^V \frac{d^{K_i}\phi_i}{K_i!} \biggr]
  \prod_{i=2}^{V} \prod_{I=1}^{K_{i-1}} \prod_{J=1}^{K_i} \frac{-(\epsilon_1+\phi^i_J+\gamma^{i,0}_J-\phi^{i-1}_I)(\epsilon_2+\phi^i_J+\gamma^{i,0}_J-\phi^{i-1}_I)}{(-\epsilon_1-\epsilon_2+\phi^{i-1}_I+\gamma^{i-1,+1}_I-\phi^i_J)(\phi^i_J+\gamma^{i,-1}_J-\phi^{i-1}_I)}
  \\
  \times \prod_{i=1}^V \biggl[ \Bigl(\frac{\epsilon_1+\epsilon_2}{\epsilon_1\epsilon_2}\Bigr)^{K_i} \prod_{1\leq I\neq J\leq K_i}
  \frac{-(-\epsilon_1-\epsilon_2+\phi^i_I+\gamma^{i,+1}_I-\phi^i_J)(\phi^i_J-\phi^i_I)} {(\epsilon_1+\phi^i_J+\gamma^{i,0}_J-\phi^i_I)(\epsilon_2+\phi^i_J+\gamma^{i,0}_J-\phi^i_I)} \biggr]
  \\
  \times \prod_{i=1}^{V} \prod_{I=1}^{K_i} \frac{\prod_{A=1}^{N_{i-1}} (\phi^i_I-a^{i-1}_A+\epsilon_1+\epsilon_2) \prod_{A=1}^{N_{i+1}} (a^{i+1}_A-\phi^i_I)} {\prod_{A=1}^{N_i} [ (\phi^i_I-a^i_A+\epsilon_1+\epsilon_2)(a^i_A-\phi^i_I) ]}
\end{multline}
for small generic shifts $\gamma^{i,0}_I$ and $\gamma^{i,\pm}_I$.  The shifts in numerators make some terms vanish more straightforwardly later on.  Eventually we will take all of these shifts to zero.

Let us prove that these shifts are generic enough to avoid higher-order poles, more precisely that at the intersection~$p_{\mathcal{D}}$ of any set~$\mathcal{D}$ of hyperplanes as above, no other denominator vanishes.  We can easily find the value of $\phi^i_I$ (and specifically what combination of~$\gamma$ appears) at the intersection~$p_{\mathcal{D}}$: consider the (unique) path $(i,I)=(i_p,I_p) \to(i_{p-1},I_{p-1})\to\cdots\to(i_1,I_1)\xrightarrow{A}0$ to the root of the tree~$\Gamma_{\mathcal{D}}$, then
\begin{equation}\label{phiintermsofgamma}
\phi^i_I = a^{i_1}_A+m\epsilon_1+n\epsilon_2+
\sum_{1\leq\alpha\leq p-1}
\gamma^{i_{\alpha},(i_{\alpha+1}-i_{\alpha})}_{I_{\alpha}}
\end{equation}
for some $m,n\in\mathbb{Z}$ that depend on the coloring of~$ \Gamma_{\mathcal{D}}$.  Since the $a^i_A$ and $a^i_A-\epsilon_1-\epsilon_2$ are generic, it is clear that $(\phi^i_I-a^i_A+\epsilon_1+\epsilon_2)$ and $(a^i_A-\phi^i_I)$ cannot vanish unless they are in~$\mathcal{D}$.  Other denominators take the form $h(\phi)=m\epsilon_1+n\epsilon_2+\phi^i_I+\gamma^{i,s}_I-\phi^{i+s}_J$ for $m,n\in\{0,1\}$ and $s\in\{0,\pm 1\}$.   Given~\eqref{phiintermsofgamma}, $\phi^{i+s}_J$ can only involve $\gamma^{i,s}_I$ if the path from $(i+s,J)$ to $0$ contains an edge $(i+s,K)\to(i,I)$, and the lack of additional $\gamma$ in $h(\phi)$ implies that $h(\phi)$ can only vanish if $K=J$, in which case $h(\phi)\in\mathcal{D}$ (this uses $\epsilon_1\neq\epsilon_2$ to deal with coloring).

The choice of shifts has a second advantage: if $\Gamma_{\mathcal{D}}$ contains two edges with the same target and the same color, then the residue at the given pole vanishes.  This is due to the numerator factor $\phi^i_I-\phi^j_J=0$ for $(i,I)$ and $(j,J)$ the sources of the two edges.  The partition function is thus a sum of residues labeled by colored trees with the restriction that no two edges share the same target and color.  Each of these residues is obtained by removing from~\eqref{tweakedz4dintegral} the factors in~$\mathcal{D}$.

We now show that in the limit $\gamma^{i,-1}_I\to0$ none of the residues are singular.  First, it is clear that $(\phi^i_I-a^i_A+\epsilon_1+\epsilon_2)$ and $(a^i_A-\phi^i_I)$ cannot vanish unless they are in~$\mathcal{D}$.  Second, $h(\phi)=\cdots+ \phi^i_I+\gamma^{i,s}_I-\phi^{i+s}_J$ for $s\in\{0,1\}$ can only vanish in the limit if there is a path $(i+s,J)\to\cdots\to(i+s,K) \to(i,I)$ where the dots only contain edges $(j-1,L)\to(j,L)$ that contribute $\gamma^{j,-1}_L\to0$.  All such edges shift $j$ in the same direction, and we deduce that there is just a single edge $(i+s,J)\to(i,I)$, which leads to $h(\phi)\in\mathcal{D}$.  Finally, $h(\phi)=\phi^i_I+\gamma^{i,-1}_I-\phi^{i-1}_J\to\phi^i_I-\phi^{i-1}_J$ can only vanish in the limit if the paths $(i,I)=(i_p,I_p)\to\cdots\to(i_1,I_1)\xrightarrow{A}0$ and $(i-1,J)=(j_q,J_q)\to\cdots\to(j_1,J_1)\xrightarrow{B}0$ only differ by edges of the form $(k-1,K)\to(k,K)$.  In particular $A=B$ and $i_1=j_1$.  Since $\Gamma_{\mathcal{D}}$ is a tree, these differing edges must be at the beginning of the two paths, hence $n$ edges are shared, $(i_m,I_m)=(j_m,J_m)$ for $1\leq m\leq n$, and the other edges obey $i_m=i_{m-1}-1$ for $n<m\leq p$ and $j_m=j_{m-1}-1$ for $n<m\leq q$.  In particular, $q=p+1$.  If $p>n$ the two edges $(i_{n+1},I_{n+1})\to(i_n,I_n)$ and $(j_{n+1},J_{n+1}) \to(j_n,J_n)$ have the same target and (no) color, but we already eliminated such trees before taking the limit $\gamma^{\dots,-1}_{ \dots}\to0$.  Thus, $p=n$ and we find an edge $(i-1,J)=(j_{n+1},J_{n+1})\to(j_n,J_n)=(i,I)$, namely $h(\phi)\in\mathcal{D}$.

The limit is smooth so we can now set all $\gamma^{i,-1}_I=0$.  Let us now show that terms for which $\mathcal{D}$ contains some $\phi^i_I-\phi^{i-1}_J$ vanish.  Among such denominators in $\mathcal{D}$ select one with maximal~$i$.  The tree $\Gamma_{\mathcal{D}}$ must contain an edge with source $(i,I)$.  If it is $(i,I)\xrightarrow{A}0$ then $\phi^i_I=a^i_A$ and the numerator $(a^i_A-\phi^{i-1}_J)$ vanishes at~$p_{\mathcal{D}}$, killing the residue.  If it is $(i,I)\xrightarrow{\epsilon_1}(i,K)$ then $\phi^i_I=\phi^i_K+\gamma^{i,0}_K+\epsilon_1$ and the numerator $\epsilon_1+ \phi^i_K+\gamma^{i,0}_K-\phi^{i-1}_J$ vanishes; similarly for $\epsilon_1$ replaced by $\epsilon_2$.  If it is $(i,I)\to(i-1,K)$ then $\phi^i_I=-\epsilon_1-\epsilon_2+\phi^{i-1}_K+ \gamma^{i-1,+1}_K$ and the numerator $-\epsilon_1-\epsilon_2+ \phi^{i-1}_K+\gamma^{i-1,+1}_K-\phi^{i-1}_J$ vanishes.  Finally, if it is $(i,I)\to(i+1,K)$ then $i$ was not maximal.

We are now left with a sum of residues labeled by colored trees $\Gamma_{\mathcal{D}}$ that do not contain pairs of edges with the same target and color and that do not contain any edge $(i-1,\dots)\to(i, \dots)$.

The absence of such edges means that the residue can be computed in several steps, first computing the residue of factors involving only $\phi^1$, then factors involving $\phi^2$ too, and so on.  Up to an ordering of each $\{\phi^i_I\mid1\leq I\leq N_i\}$, which cancels the $1/K_i!$ factor, the partition function is thus obtained by taking the residue for $\phi^i_I$ in lexicographic order $\phi^1_1, \phi^1_2,\ldots,\phi^2_1,\phi^2_2,\ldots$ where at step $\phi^i_I$ we only select poles from denominators $(\cdots-\phi^i_I)$ that do not depend on components $\phi^j_J$ with $(j,J)>(i,I)$ lexicographically.  We can now take the limit $\gamma \to0$ while keeping this selection of poles.

Let us prove that the poles that remain in this limit follow Nekrasov's prescription~\eqref{expectedsetofpoles}.  We assume for some $1\leq U\leq V$ that
\begin{equation}\label{inductionpoles1}
\{\phi^i_I\mid1\leq I\leq K_i\} = \{a^i_A+(r-1)\epsilon_1+(s-1)\epsilon_2\mid1\leq A\leq N_i,(r,s)\in\lambda^i_A
\}
\end{equation}
for all $1\leq i<U$ and we assume for some $1\leq K\leq K_U$ that
\begin{equation}\label{inductionpoles2}
\{\phi^U_I\mid1\leq I<K\} = \{a^U_A+(r-1)\epsilon_1+(s-1)
\epsilon_2\mid1\leq A\leq N_U,(r,s)\in\mu_A\}
\end{equation}
for some Young diagrams $\lambda^i_A$ and $\mu_A$ (the latter depend on $U$ and $K$ and are eventually contained in $\lambda^U_A$).  From these induction hypotheses we show that the next component, $\phi^U_K$, takes a value that corresponds to adding a box to one of the $\mu_A$.  To do this, we compute the partial residue at \eqref{inductionpoles1} and \eqref{inductionpoles2} and focus on factors that depend on $\phi^U_K$ and no higher $\phi^i_I$:
\begin{multline}
  \prod_{A=1}^{N_U} \frac{\prod_{(r,s)\in\partial_-\mu_A}[(\phi^U_K-a^U_A-(r-1)\epsilon_1-(s-1)\epsilon_2)(a^U_A+r\epsilon_1+s\epsilon_2-\phi^U_K)]} {\prod_{(r,s)\in\partial_+\mu_A}[(\phi^U_K-a^U_A-(r-2)\epsilon_1-(s-2)\epsilon_2)(a^U_A+(r-1)\epsilon_1+(s-1)\epsilon_2-\phi^U_K)]}
  \\
  \times \prod_{A=1}^{N_{U+1}} (a^{U+1}_A-\phi^U_K) \prod_{A=1}^{N_{U-1}} \frac{\prod_{(r,s)\in\partial_+\lambda^{U-1}_A}(\phi^U_K-a^{U-1}_A-(r-2)\epsilon_1-(s-2)\epsilon_2)} {\prod_{(r,s)\in\partial_-\lambda^{U-1}_A}(\phi^U_K-a^{U-1}_A-(r-1)\epsilon_1-(s-1)\epsilon_2)} \,.
\end{multline}

Here, $\partial_-\mu\subset\mu$ (respectively $\partial_+ \mu$) denote the sets of boxes that could be removed from (respectively added to) $\mu$ in such a way that the result is still a Yong diagram.  The prescription selects poles from denominators of the form $(\cdots-\phi^U_K)$ hence poles at $\phi^U_K=a^U_A+(r-1) \epsilon_1+(s-1)\epsilon_2$ for $(r,s)\in\partial_+\mu_A$.  This concludes the induction and the overall proof.

While in this appendix the poles of the bifundamental contributions $S(\phi^{i+1}_J-\phi^i_I)$ did not contribute, we study in the main text some other JK prescriptions for which such poles contribute.

\section{2d \texorpdfstring{$\mathcal{N}=(2,2)$}{N=(2,2)} partition functions}\label{b:2d}

We now review vortex partition functions of 2d $\Nsusy=(2,2)$ gauge theories (with vector and chiral multiplets).  They can be extracted from the partition function on a sphere~$S^2$ of radius $1/\epsilon$ \cite{Benini:2012ui,2dvortex}, recently reviewed in~\cite{Benini:2016qnm,Park:2016dpb}.

We start with the Coulomb representation of the sphere partition function:
\begin{equation}\label{zs2}\compressthisequation
  Z_{S^2} = \frac{1}{|\mathcal{W}|} \sum_{B\in\mathfrak{t}_{\text{GNO}}} \int_{\mathfrak{t}} d\hat{\sigma} \,
  \prod_{\ell} \Bigl[ z_{\ell}^{\Tr_{\ell} i\hat{\sigma}^+} \bar{z}_{\ell}^{\Tr_{\ell} i\hat{\sigma}^-} \Bigr]
  \prod_{\alpha\in\text{roots}(G)} \!\!\frac{\Gamma(1-\alpha\cdot i\hat{\sigma}^+)}{\Gamma(\alpha\cdot i\hat{\sigma}^-)}
  \prod_{w\in\text{weights}}\frac{\Gamma(-w\cdot i\hat{\sigma}^+)}{\Gamma(1+w\cdot i\hat{\sigma}^-)} \,.
\end{equation}
Here $\hat{\sigma}=\sigma/\epsilon$ is the dimensionless Coulomb branch scalar and $i\hat{\sigma}^{\pm}=i\hat{\sigma}\pm B/2$ in terms of the magnetic flux~$B$.  The integral ranges over the Cartan algebra~$\mathfrak{t}$ of the gauge group~$G$, and the sum ranges over a lattice in~$\mathfrak{t}$ defined by GNO quantization: $B$~has integer eigenvalues on any representation of~$G$ (this depends on the global structure of~$G$).  Dividing by the cardinal of the Weyl group cancels a gauge redundancy.  For each $U(1)$ gauge factor the theory has an FI parameter~$\xi_{\ell}$ and a theta angle $\theta_{\ell}$ which combine into $\tau_{\ell}=\theta_{\ell}/(2\pi)+i\xi_{\ell}$ or equivalently $z_{\ell}=e^{2\pi i\tau_{\ell}}$.  The other products range over roots $\alpha$ of the gauge group~$G$ and weights~$w$ (with multiplicity) of the representation~$\mathcal{R}$ of~$G$ under which chiral multiplets transform.  We expressed the vector multiplet one-loop determinant (product over $\alpha$) in an unconventional way but it is equal to the usual form because
\begin{equation}
\prod_{\alpha>0} \biggl[ \frac{\Gamma(1-\alpha\cdot i\hat{\sigma}^+)}{\Gamma(\alpha\cdot i\hat{\sigma}^-)}\frac{\Gamma(1+
\alpha\cdot i\hat{\sigma}^+)}{\Gamma(-\alpha\cdot i
\hat{\sigma}^-)} \biggr] = \prod_{\alpha>0} \biggl[ (\alpha
\cdot i\hat{\sigma}^+)(-\alpha\cdot i\hat{\sigma}^-) \frac{
\sin\pi(\alpha\cdot\hat{\sigma}^-)}{\sin\pi(\alpha\cdot
\hat{\sigma}^+)} \biggr]
\end{equation}
using $\Gamma(x)\Gamma(1-x)=\pi/\sin\pi x$, and the ratio of sines gives the expected sign $(-1)^{\alpha\cdot B}$.
Each chiral multiplet, namely each irreducible component of~$\mathcal{R}$, can be given a (vector) $R$-charge~$r$ (we assume $0<r<2$) and a twisted mass~$m$.  On the sphere this turns out to be equivalent to giving a VEV $ \hat{\sigma}=m/\epsilon+ir/2$ (henceforth called the complexified twisted mass) for a background vector multiplet scalar~$\hat{\sigma}$ coupled to that chiral multiplet's flavor symmetry.  We thus omit $R$-charges and twisted masses from notations.  Finally, the theory can have a superpotential, but its only effect on $Z_{S^2}$ is to break flavor symmetries by constraining some linear combinations of complexified twisted masses (hence of $R$-charges).

We will assume that all poles of the integrand in~\eqref{zs2} are simple.  This holds for all quiver gauge theories we consider in this paper as long as twisted masses are generic, however it does not hold for instance in a theory where a $U(k)$ gauge group has two adjoint chiral multiplets, even if their masses are generic.  When the theory has a superpotential, twisted masses cannot be assumed to be generic and one then encounters higher-order poles.  These can often be resolved by first ignoring the superpotential constraint on complexified twisted masses then taking the limit where the constraint is obeyed.

The integrand in~\eqref{zs2} has lattices of poles for which $w\cdot i\hat{\sigma}$ differ by integers; these lattices are (or can be further subdivided into) cones of $\mathfrak{t}_{\text{GNO}}$ shifted by some constant $i\hat{\sigma}_0$.  Under some convergence assumptions, the contour integral can be evaluated as a sum of residues at some of these lattices of poles.  Thanks to the factorization of the integrand into a function of $z$ and $i\hat{\sigma}^+$ times a function of $\bar{z}$ and $i\hat{\sigma}^-$, the $S^2$~partition function becomes
\begin{equation}\label{appzs2higgsbranch}
Z_{S^2} = \sum_{\text{some}\ i\hat{\sigma}_0} Z^{\text{2d}}_{
\text{pert}}(i\hat{\sigma}_0;z\bar{z}) Z^{\text{2d}}_{
\text{vortex}}(i\hat{\sigma}_0;z) Z^{\text{2d}}_{\text{vortex}}(i
\hat{\sigma}_0;\tilde{\bar{z}})
\end{equation}
with $\tilde{z}=z(-1)^{\sum w}$ namely $\prod_{\ell}\tilde{z}_{\ell}^{\Tr_{\ell}i\hat{\sigma}}=\prod_{\ell}z_{\ell}^{\Tr_{\ell}i\hat{\sigma}}\prod_w e^{i\pi w \cdot i\hat{\sigma}}$, and $\tilde{\tilde{\bar{z}}}=\bar{z}$ and
\begin{align}
  Z^{\text{2d}}_{\text{pert}}(i\hat{\sigma}_0;z\bar{z}) & =
  \prod_{\ell} (z_{\ell} \bar{z}_{\ell})^{\Tr_{\ell} i\hat{\sigma}_0}
  \prod_{\alpha\in\text{roots}(G)} \frac{\pi}{\sin\pi(\alpha\cdot i\hat{\sigma}_0)}
  \prod_{w\in\text{weights}}\frac{\pi}{\sin\pi(1+w\cdot i\hat{\sigma}_0)} \,,
  \\\label{appzvassum}
  Z^{\text{2d}}_{\text{vortex}}(i\hat{\sigma}_0;z) & = \!\!\!\sum_{k\in\mathfrak{t}_{\text{GNO}}} \biggl[ \prod_{\ell} \tilde{z}_{\ell}^{\Tr_{\ell}(k)} \!\!\!\prod_{\alpha\in\text{roots}(G)} \frac{1} {\Gamma(\alpha\cdot i(\hat{\sigma}_0+k))}\prod_{w\in\text{weights}}\frac{1}{\Gamma(1+w\cdot i(\hat{\sigma}_0+k))} \biggr] \,.
\end{align}
While the sum ranges over a lattice one should remember that the Gamma functions we started with only have a semi-infinite set of poles.  This implies that the terms here are only non-zero for $k$ in some cone inside $\mathfrak{t}_{\text{GNO}}$.  The convergence assumption mentioned above is that the remaining series should converge for each $i\hat{\sigma}_0$ appearing in~\eqref{appzs2higgsbranch}.  This holds for all cases we consider in the main text, but it requires the gauge theory to have enough (generic) FI parameters.

The value of~$\hat{\sigma}_0$ must be chosen at the root of a Higgs branch: the finitely many allowed values can be found\footnote{The three procedures are expected to be equivalent for generic twisted masses.  This has not been proven directly yet, but rather by requiring that different localization computations agree.} from the $S^2$~partition function, or by solving $D$-term equations of the theory, or by requiring that all terms vanish for~$k$ outside a cone (not necessarily based at the origin).  This last condition helps ensure that there exists a range of FI parameters $\log\lvert z\rvert$ for which the sum converges.

Up to a normalization and a shift in masses (this is a matter of conventions), $Z^{\text{2d}}_{\text{vortex}}$ is the vortex partition function.  It can be rewritten as an integral, again up to a normalization,
\begin{equation}\label{appzvasint}
  Z^{\text{2d}}_{\text{vortex}}(i\hat{\sigma}_0;z) = \int_{\gamma(i\hat{\sigma}_0)} d\hat{\sigma} \, \prod_{\ell} z_{\ell}^{\Tr_{\ell}i\hat{\sigma}} \frac{\prod_{w\in\text{weights}}\Gamma(-w\cdot i\hat{\sigma})}{\prod_{\alpha\in\text{roots}(G)}\Gamma(\alpha\cdot i\hat{\sigma})}
\end{equation}
but the contour $\gamma(i\hat{\sigma}_0)$ is hard to describe except by the list of poles it encloses: $i\hat{\sigma}\in i\hat{\sigma}_0+\mathfrak{t}_{\text{GNO}}$.
Note that \eqref{appzvassum} and \eqref{appzvasint} differ by $1/\Gamma(1+w\cdot i\hat{\sigma})\to\Gamma(-w\cdot i\hat{\sigma})$ in other words by $\sin\pi(-w\cdot i\hat{\sigma})/\pi$.  This sine takes the same value at all poles $i\hat{\sigma}_0+\mathfrak{t}_{\text{GNO}}$ hence only changes the normalization, except for a sign.  This sign is absorbed by changing $z\leftrightarrow\tilde{z}$.  The other effect of the sine is to go from an expression~\eqref{appzvasint} with many lattices of poles to one with no pole~\eqref{appzvassum}.

As a final comment let us mention that the $S^2$~partition function in the presence of 4d instantons at the North and South poles is obtained by including $\prod_{\pm} Z^{\text{4d--2d}}(\phi^{\pm},i\sigma^{\pm})$ in~\eqref{zs2}.  As emphasized in the main text, we also get non-zero residues from some negative vortex numbers~$k$.

\printbibliography
\end{document}